\documentclass[aps,prd,preprintnumbers,groupedaddress,nofootinbib,twocolumn,amssymb,eqsecnum,nofootinbib,notitlepage]{revtex4-1}
\usepackage[dvipdfmx]{graphicx}
\usepackage{bm}
\usepackage{amsmath}
\usepackage{color}
\usepackage{hyperref}
\usepackage{amsfonts}
\usepackage[subnum]{cases}
\usepackage{ulem}

\begin{document}
\newcommand{\newc}{\newcommand}

\newc{\be}{\begin{equation}}
\newc{\ee}{\end{equation}}
\newc{\ba}{\begin{eqnarray}}
\newc{\ea}{\end{eqnarray}}
\newc{\bea}{\begin{eqnarray*}}
\newc{\eea}{\end{eqnarray*}}
\newc{\D}{\partial}
\newc{\ie}{{\it i.e.} }
\newc{\eg}{{\it e.g.} }
\newc{\etc}{{\it etc.} }
\newc{\etal}{{\it et al.}}
\newc{\Mpl}{M_{\rm Pl}}
\newcommand{\nn}{\nonumber}
\newc{\ra}{\rightarrow}
\newc{\lra}{\leftrightarrow}
\newc{\lsim}{\buildrel{<}\over{\sim}}
\newc{\gsim}{\buildrel{>}\over{\sim}}
\def\mpl{M_{\rm pl}}
\def\d{\mathrm{d}}

\newcommand{\SP}{\color{blue}}

\preprint{WUCG-22-10}
\title{Cosmology in theories with spontaneous scalarization of neutron stars}

\author{
Ratchaphat Nakarachinda$^{1}$, 
Sirachak Panpanich$^{2}$, 
Shinji Tsujikawa$^{2}$, and 
Pitayuth Wongjun$^{1}$}

\affiliation{
$^1$The Institute for Fundamental Study,
Naresuan University, Phitsanulok, 65000, Thailand\\
$^2$Department of Physics, Waseda University, 3-4-1 Okubo, Shinjuku, Tokyo 169-8555, Japan
}
\date{\today}

\begin{abstract}

In a model of spontaneous scalarization of neutron stars proposed 
by Damour and Esposite-Farese, a general relativistic branch becomes 
unstable to trigger tachyonic growth of a scalar field $\phi$ toward a 
scalarized branch. Applying this scenario to cosmology, there is fatal tachyonic instability of $\phi$ during inflation and 
matter dominance being incompatible with solar-system constraints on today's field value 
$\phi_0$. In the presence of a four-point coupling
$g^2 \phi^2 \chi^2/2$ between $\phi$ and an inflaton field $\chi$, it was argued by Anson {\it et al.} that a positive mass squared heavier 
than the square of a Hubble expansion rate leads to the exponential 
suppression of $\phi$ during inflation and that $\phi_0$
can remain small even with the growth of $\phi$ after the radiation-dominated epoch.
For several inflaton potentials approximated as 
$V(\chi)=m^2 \chi^2/2$ about the potential minimum, we study the dynamics 
of $\phi$ during reheating as well as other cosmological epochs in detail.
For certain ranges of the coupling $g$, the homogeneous field $\phi$ can be 
amplified by parametric resonance during a coherent oscillation of the inflaton.
Incorporating the backreaction of created particles under a Hartree approximation, 
the maximum values of $\phi$ reached during preheating are 
significantly smaller than those obtained without the backreaction.
We also find that the minimum values of $g$ consistent with solar system 
bounds on $\phi$ at the end of reheating are of order $10^{-5}$ and hence 
there is a wide range of acceptable values of $g$.
Thus, the scenario proposed by Anson {\it et al.} naturally leads to the viable cosmological evolution of $\phi$ consistent with local gravity constraints, 
without modifying the property of scalarized neutron stars.

\end{abstract}

\maketitle

\section{Introduction}

The physics in strong gravity regimes can be now probed by 
the observations of gravitational waves (GWs) emitted from a binary 
system containing black holes or neutron stars (NSs) \cite{LIGOScientific:2016aoc,LIGOScientific:2017vwq,LIGOScientific:2020aai}. 
With the future high precision data of GWs, it will be possible to 
probe the accuracy of General Relativity (GR) and the possible 
deviation from it \cite{Berti:2015itd,Barack:2018yly,Barack:2018yly,Berti:2018cxi}. 
In theories beyond GR, there are usually additional 
degrees of freedom that can be coupled to 
gravity \cite{Copeland:2006wr,DeFelice:2010aj,Clifton:2011jh,Joyce:2014kja,Heisenberg:2018vsk,Kase:2018aps,Baker:2019gxo}. 
A simplest example is known as scalar-tensor 
theories, in which a scalar field $\phi$ has nonminimal 
or derivative couplings to gravity \cite{Brans:1961sx,Fujii:1982ms,Fujii:2003pa,Horndeski:1974wa,Deffayet:2011gz,Kobayashi:2011nu,Charmousis:2011bf}.

In the presence of a nonminimal coupling $F(\phi)R$, where 
$F$ is a function of $\phi$ and $R$ is the Ricci scalar, 
it is known that NSs can have scalar hairs through 
an indirect coupling between the scalar field and matter 
mediated by gravity \cite{Damour:1993hw,Damour:1996ke,Harada:1998ge,
Novak:1998rk,Sotani:2004rq,Cooney:2009rr,Arapoglu:2010rz,Orellana:2013gn,AparicioResco:2016xcm,Kase:2019dqc}. 
For the exponential coupling 
$F(\phi)=e^{-2 Q\phi/\Mpl}$, where $Q$ is a constant 
and $\Mpl$ is the reduced Planck mass, the fifth force 
propagates around weak gravitational objects like the Sun. 
From solar-system experiments, there is a tight bound 
$|Q|<2 \times 10^{-3}$ on the coupling
strength \cite{Will:2014kxa,Tsujikawa:2008uc}. 
Then, the deviation from GR in the vicinity of NSs is 
also suppressed, so probing the modification of gravity 
from the GW observations is challenging for the nonminimal 
coupling $F(\phi)=e^{-2 Q\phi/\Mpl}$.

If we consider nonminimal couplings containing even power-law functions 
of $\phi$, it is possible to have a nontrivial branch  
$\phi'(r) \neq 0$ besides a GR branch $\phi'(r)=0$
on a static and spherically symmetric background 
with the radial distance $r$. A typical example is the coupling 
$F(\phi)=e^{-\beta \phi^2/(2\Mpl^2)}$ proposed by 
Damour and Esposite-Farese (DEF), 
where $\beta$ is a constant \cite{Damour:1993hw,Damour:1996ke}. 
For strong gravitational objects like NSs, 
the GR branch can be unstable to trigger tachyonic instability of 
the scalar field to reach a nontrivial branch of $\phi$. 
This phenomenon, which is called spontaneous scalarization, 
occurs for negative coupling constants in the range 
$\beta \leq -4.35$ \cite{Harada:1998ge,Novak:1998rk,Silva:2014fca,Barausse:2012da}. 
Since the gravitational interaction for such scalarized NSs exhibits 
the appreciable deviation from that in GR, it is possible to probe 
the modification of gravity from binary pulsar 
measurements \cite{Freire:2012mg,Shao:2017gwu,Anderson:2019eay} 
as well as the observations of GWs emitted from a binary system 
containing at least one NS \cite{Niu:2021nic,Higashino:2022izi}.

When spontaneous scalarization of NSs occurs, the asymptotic value 
of the scalar field $\phi_0$ needs to be in the range 
$|\phi_0| \lesssim 10^{-3}\Mpl |\beta|^{-1}$ for the 
consistency with local gravity constraints \cite{Damour:1993hw}. 
This asymptotic value should be determined by the cosmological 
evolution of $\phi$ from the past to today. 
In the original DEF model, however, the background cosmological 
scalar field largely deviates from 0 for the coupling range $\beta$
allowing for spontaneous scalarization \cite{Damour:1992kf,Anderson:2016aoi}. 
This is attributed to the fact that the negative coupling $\beta$ 
leads to tachyonic growth of $|\phi|$ during the cosmological 
evolution. Without severely fine-tuned initial conditions, 
the amplitude of today's field value exceeds the upper limit 
constrained by solar system tests of gravity. 
We note that the similar type of instabilities is also present  
for spontaneously scalarized BHs arising from a scalar 
Gauss-Bonnet coupling \cite{Anson:2019uto,Franchini:2019npi,Antoniou:2020nax}.

On the other hand, Anson {\it et al.} \cite{Anson:2019ebp} proposed a mechanism 
of reconciling spontaneous scalarization with cosmology 
by incorporating a coupling $g^2 \phi^2 \chi^2/2$ between the 
scalar field $\phi$ and an inflaton field $\chi$.
Since the field $\phi$ can have a positive effective mass squared 
$g^2 \chi^2$ larger than the squared Hubble expansion rate 
during inflation, the amplitude of the homogeneous field $\phi$ 
exponentially decreases. 
Even though $|\phi|$ increases after the onset of the radiation 
era to today, it is expected that today's field value is still in the range 
$|\phi_0| \lesssim 10^{-3}\Mpl |\beta|^{-1}$. 
However, the analysis of Ref.~\cite{Anson:2019ebp} does not accommodate 
the evolution of $\phi$ during the post inflationary reheating period.
Indeed, for certain ranges of the coupling $g$, the four-point coupling 
$g^2 \phi^2 \chi^2/2$ leads to 
parametric resonance of the homogeneous field $\phi$ and its perturbations 
during a preheating stage after inflation \cite{Kofman:1994rk,Shtanov:1994ce,Kaiser:1995fb,Khlebnikov:1996zt,Khlebnikov:1996wr,Prokopec:1996rr,Kofman:1997yn,Bassett:2005xm}. Even for small couplings $g$ without the period of preheating, 
it can happen that the amplitude of $\phi$ grows during reheating 
by the dominance of the negative nonminimal coupling over 
the positive mass term $g^2 \chi^2$. 
Hence it is important to clarify the coupling range of $g$ in which 
the model can be at work.
We note that there are other mechanisms for reconciling spontaneous 
scalarization with cosmology \cite{Anderson:2016aoi,Silva:2019rle,Minamitsuji:2022qku,Higashino:2022izi}, but it is typically nontrivial to realize the acceptable cosmological evolution of $\phi$.

In the DEF model with the coupling $g^2 \phi^2 \chi^2/2$, we will study 
the cosmological evolution of the scalar field $\phi$ responsible 
for spontaneous scalarization. We pay particular attention to 
the dynamics during reheating in which the further growth of 
$|\phi|$ can be expected. 
Since the inflaton potentials are approximated as $V(\chi) \simeq m^2 \chi^2/2$ 
around $\chi=0$, the reheating period corresponds to 
an effective matter era driven by the oscillation of a 
massive inflaton field.

For coupling ranges of $g$ in which the preheating stage is present, 
we need to take into account the backreaction of created $\phi$ particles 
that leads to the violation of coherent oscillations of $\chi$. 
Without the backreaction, the maximum amplitude 
$\phi_{\rm max}$ of the field $\phi$ reached during preheating can exceed a value 
constrained by solar-system tests of gravity at the end of reheating 
($|\phi_{\rm R}| \lesssim 10^{-11}\Mpl$). 
However, we will show that implementing the backreaction
under the Hartree approximation \cite{Khlebnikov:1996wr,Kofman:1997yn} 
leads to $\phi_{\rm max}$ significantly smaller than $10^{-11}\Mpl$. 
For two inflaton potentials considered in this paper, 
$\phi_{\rm max}$ is less than the order of $10^{-38}\Mpl$.
After the system reaches an equilibrium state with the violation 
of coherent oscillations of the inflaton, the further significant 
amplification of $\phi$ is not expected by the end of reheating 
because the negative nonminimal coupling is suppressed compared 
to $g^2 \chi^2$ in the background equation of $\phi$.
Thus, even with the preheating epoch, the presence of the coupling $g^2 \phi^2 \chi^2/2$ allows the cosmological 
evolution of 
$\phi$ consistent with local gravity constraints on $\phi_0$. 

For small coupling ranges of $g$ in which preheating does not occur, 
we do not need to implement the backreaction of created particles. 
In such cases, we will solve the background equations of motion 
by the end of reheating with a Born decay constant $\Gamma$ 
taken into account.
Depending on the form of inflaton potentials, the nonminimal coupling 
can overwhelm the term $g^2 \chi^2$ during reheating.  
Since this leads to the growth of $|\phi|$ by the end of reheating, 
it can happen that $|\phi_{\rm R}|$ exceeds the upper bound $10^{-11}\Mpl$. 
This is especially the case for a low-scale reheating scenario 
with the reheating temperature of order MeV.
For two inflaton potentials, we will put lower bounds on the coupling 
$g$ consistent with solar-system constraints. 
In both cases, the minimum values of $g$ are of order $10^{-5}$, 
so the mechanism proposed by Anson {\it et al.} \cite{Anson:2019ebp} 
is at work for wide ranges of the coupling $g$ (including the 
case where preheating occurs).

This paper is organized as follows.
In Sec.~\ref{scasec}, we briefly review the DEF model and derive 
the background equations of motion on the spatially-flat 
Friedmann-Lema\^{i}tre-Robertson-Walker (FLRW) 
spacetime in the presence of the coupling $g^2 \phi^2 \chi^2/2$.
In Sec.~\ref{latesec}, we discuss the cosmological evolution of $\phi$ 
from the radiation era to today and interpret a solar-system 
bound on $\phi_0$ as the constraint on $\phi$ 
at the onset of radiation dominance.
In Sec.~\ref{infsec}, we study the dynamics of the scalar field 
during inflation for several inflaton potentials and find minimum values 
of $g$ above which the amplitude of $\phi$ exponentially decreases. 
In Sec.~\ref{reheatsec}, we analyze the evolution of the homogeneous 
field $\phi$ and its perturbations during the reheating stage and 
derive minimum values of the coupling constant $g$ consistent 
with solar-system constraints.
Sec.~\ref{consec} is devoted to conclusions.

\section{Theories with spontaneous scalarization}
\label{scasec}

We consider theories given by the action 
\ba
{\cal S} &=& 
\int {\rm d}^4 x \sqrt{-g_J} 
\left[ \frac{\Mpl^2}{2}F(\phi) R+\omega (\phi)X
+L_{\rm inf} \right]
\nonumber\\
& &
+{\cal S}_m (g_{\mu \nu}, \Psi_m)\,,
\label{action}
\ea
where $g_J$ is a determinant of metric tensor $g_{\mu \nu}$ in the 
Jordan frame, 
and $X=-(1/2)g^{\mu \nu} \nabla_{\mu} \phi \nabla_{\nu} \phi$ is 
a scalar kinetic term with $\nabla_{\mu}$ being a covariant 
derivative operator. 
The $\phi$-dependent function in front 
of $X$ is chosen to be \cite{Kase:2020yhw,Kase:2020qvz,Higashino:2022izi}
\be
\omega(\phi)=\left( 1-\frac{3\Mpl^2 F_{,\phi}^2}{2F^2} 
\right)F(\phi)\,,
\label{omega}
\ee
where $F_{,\phi}={\rm d}F/{\rm d}\phi$. 
Brans-Dicke (BD) theories \cite{Brans:1961sx} correspond to the particular 
nonminimal coupling $F(\phi)=e^{-2Q \phi/\Mpl}$, where a coupling 
constant $Q$ is related 
to the BD parameter 
$\omega_{\rm BD}$ as $3+2\omega_{\rm BD}=1/(2Q^2)$ \cite{Tsujikawa:2008uc,Khoury:2003rn}. 
In theories of spontaneous scalarization, $F(\phi)$ contains 
even power-law functions of $\phi$.

We take into account the contribution of an inflaton field $\chi$ 
as the Lagrangian  
\be
L_{\rm inf}=-\frac{1}{2}g^{\mu \nu} \nabla_{\mu} \chi 
\nabla_{\nu} \chi
-V(\chi)-\frac{1}{2}g^2 \phi^2 \chi^2\,,
\label{Linf}
\ee
where $V$ is the potential of $\chi$. 
The last term in Eq.~(\ref{Linf}) characterizes an interaction between 
$\phi$ and $\chi$ with a coupling constant $g$. 
During inflation, this can generate a large effective 
positive mass squared of $\phi$ relative to the square of 
the Hubble expansion rate.
Then, it can compensate a negative mass squared 
induced by the nonminimal coupling $F(\phi) R$ 
responsible for spontaneous scalarization. 
This allows a possibility for avoiding the tachyonic 
growth of $\phi$ during inflation \cite{Anson:2019ebp}.

In the presence of the four-point coupling $g^2 \phi^2 \chi^2/2$, 
it is known that a phenomenon called preheating \cite{Kofman:1994rk,Kofman:1997yn}
can occur after inflation during the coherent oscillation of inflaton.
In this stage, the scalar field $\phi$ and its perturbations $\delta \phi$ 
can be amplified by parametric resonance.
Since the dynamics of the field $\phi$ during preheating were not 
addressed in Ref.~\cite{Anson:2019ebp}, we will study whether 
the presence of this stage is harmless or not for the compatibility 
with local gravity constraints on today's field value $\phi_0$. 
For this purpose, we also scrutinize the scalar-field dynamics 
in other cosmological epochs from the onset of inflation 
to today.

In Eq.~(\ref{action}), the action ${\cal S}_m$ incorporates the 
contributions of matter fields $\Psi_m$ such as radiation, 
nonrelativistic matter, and dark energy. 
We assume that matter fields are minimally coupled to gravity.

Under the conformal transformation 
$\hat{g}_{\mu \nu}=F(\phi)g_{\mu \nu}$, 
the action (\ref{action}) is transformed 
to \cite{DeFelice:2010aj,Kase:2020yhw}
\ba
\hat{{\cal S}} &=&
\int {\rm d}^4 x \sqrt{-\hat{g}} \left[ \frac{\Mpl^2}{2} 
\hat{R}-\frac12 \hat{g}^{\mu \nu} \nabla_{\mu} \phi
\nabla_{\nu} \phi+\hat{L}_{\rm inf} \right] \nonumber \\
&&+{\cal S}_m (F^{-1}(\phi) \hat{g}_{\mu \nu}, \Psi_m)\,,
\ea
where a hat represents quantities in the Einstein frame.
In the transformed frame, the scalar field $\phi$ is 
coupled to matter fields through the metric tensor 
$\hat{g}_{\mu \nu}$. 

We deal with the Jordan frame as a physical one and perform 
all analyses in this frame by 
exploiting the action (\ref{action}). 
We consider the nonminimal coupling chosen 
by DEF \cite{Damour:1993hw,Damour:1996ke}
\be
F(\phi)=e^{-\beta \phi^2/
(2\Mpl^2)}\,,
\label{Fnon}
\ee
where $\beta$ is a constant. 
In this case, we have
\be
\omega(\phi)=\left( 1- \frac{3 \beta^2 \phi^2}{2 \Mpl^2} 
\right) e^{-\beta \phi^2/
(2\Mpl^2)}\,.
\ee
Spontaneous scalarization of NSs can occur for the coupling $\beta \le -4.35$ \cite{Harada:1998ge,Novak:1998rk,Silva:2014fca,Barausse:2012da}. 
In such cases, the GR branch $\phi=0$ can be unstable to trigger 
tachyonic instability toward the other nontrivial branch $\phi \neq 0$. 
From binary pulsar measurements of the energy loss through dipolar radiation, the coupling $\beta$ was constrained 
to be $\beta \ge -4.5$ \cite{Freire:2012mg,Shao:2017gwu}.   
Thus, the coupling constant $\beta$ is restricted in a limited range.

We consider a spatially-flat FLRW background given by 
the line element
\be
{\rm d}s^2=-{\rm d}t^2+a^2(t) \delta_{ij}
{\rm d}x^i {\rm d}x^j\,,
\ee
where $a(t)$ is a time-dependent scale factor.  
Incorporating the Born decay term $\Gamma \dot{\chi}$
to complete the reheating process ($\Gamma$ is 
a decay constant and a dot represents 
a derivative with respect to $t$) into 
the inflaton equation of motion, it follows that 
\be
\ddot{\chi}+\left( 3H +\Gamma \right) \dot{\chi}
+V_{,\chi}+g^2 \phi^2 \chi=0\,,
\label{chieq}
\ee
where $H=\dot{a}/a$ is the Hubble expansion rate.
We will consider the case $H \gg \Gamma$ during inflation, 
so the decay term $\Gamma \dot{\chi}$ is important only 
at the late stage of reheating.
Due to the energy transfer from the inflaton to radiation, 
the radiation energy density $\rho_r$ satisfies
the differential equation 
\be
\dot{\rho}_r+4H \rho_r=\Gamma \dot{\chi}^2\,.
\ee
The energy density $\rho_m$ of nonrelativistic 
matter (cold dark matter and baryons) obeys the 
continuity equation 
\be
\dot{\rho}_m+3H \rho_m=0\,, 
\ee
with a vanishing pressure.
As a source for dark energy, we take
the cosmological constant $\Lambda$ 
into account.
 
The (00) and (11) components of gravitational field equations following from 
the action (\ref{action}) are 
given, respectively, by 
\ba
& & 3\Mpl^2 H (FH+ F_{,\phi}\dot{\phi})=
\frac{1}{2} \dot{\chi}^2+V+\frac{1}{2} \omega \dot{\phi}^2
+\frac{1}{2} g^2 \phi^2 \chi^2
\nonumber \\
& &
+\rho_r+\rho_m+\Lambda\,,\label{back1}\\
& & -\Mpl^2 F(2\dot{H}+3H^2)
=\Mpl^2 \left( F_{,\phi} \ddot{\phi}+F_{,\phi \phi} \dot{\phi}^2
+2F_{,\phi}H \dot{\phi} \right)\nonumber \\
& &
+\frac{1}{2} \dot{\chi}^2-V+\frac{1}{2} \omega \dot{\phi}^2
-\frac{1}{2} g^2 \phi^2 \chi^2 
+\frac{1}{3}\rho_r-\Lambda\,.\label{back2}
\ea
The scalar field $\phi$ obeys the differential equation 
\be
\ddot{\phi}+3H \dot{\phi} 
+m_{\rm eff}^2\phi=0\,,
\label{phieq}
\ee
where
\be
m_{\rm eff}^2 \equiv
\frac{1}{\omega}
\left[ g^2 \chi^2+3 \beta F \left( 2H^2+\dot{H} 
\right)-\frac{\beta \omega \dot{\phi}^2}{2\Mpl^2}
-\frac{3\beta^2F \dot{\phi}^2}{2\Mpl^2} 
 \right].
\label{meff}
\ee

To discuss the dynamics of inflation and reheating, we consider 
the inflaton potentials of $\alpha$-attractors \cite{Kallosh:2013yoa} given by 
\be
V(\chi)=\frac{3}{4}\alpha m^2 \Mpl^2 
\left[ 1-\exp \left( -\sqrt{\frac{2}{3\alpha}} \frac{\chi}{\Mpl} 
\right) \right]^2\,,
\label{Vchi}
\ee
where $\alpha$ is a positive dimensionless constant, and 
$m$ is a constant having a dimension 
of mass. 
In the limit $\alpha \to \infty$, the potential (\ref{Vchi}) reduces 
to that in chaotic inflation, i.e., 
$V(\chi)=m^2 \chi^2/2$ \cite{Linde:1983gd}. 
Starobinsky's model with the Lagrangian 
$L=R+R^2/(6m^2)$ \cite{Starobinsky:1980te} 
gives rise to the potential (\ref{Vchi}) with $\alpha=1$
after a conformal transformation to the 
Einstein frame \cite{DeFelice:2010aj}.
For the numerical simulation performed in Secs.~\ref{infsec}
and \ref{reheatsec}, we will consider the two 
potentials: (i) $V(\chi)=m^2 \chi^2/2$, 
and (ii) the potential (\ref{Vchi}) with $\alpha=1$.

\section{Cosmological dynamics from radiation domination 
to today}
\label{latesec}

In this section, we investigate a bound of the field value $\phi$
at the onset of radiation dominance constrained from solar 
system tests of gravity.
In theories given by the action (\ref{action}), the post-Newtonian parameter 
$\gamma_{\rm PPN}$ is \cite{Damour:1992we,Damour:1993hw}
\be
\gamma_{\rm PPN}-1 = -\frac{2\alpha_{\rm PPN}^2 (\varphi_0)}
{1+\alpha_{\rm PPN}^2 (\varphi_0)}
=-\frac{2 \beta^2 \varphi_0^2}{1+\beta^2 \varphi_0^2}\,,
\ee
where $\varphi=\phi/(\sqrt{2}\Mpl)$ is a dimensionless field 
with today's value $\varphi_0$,   
and $\alpha_{\rm PPN}=-F_{,\varphi}/(2F)=\beta \varphi$ 
with $F=e^{-\beta \varphi^2}$.
The Shapiro time delay measurements have given the bound 
$\gamma_{\rm PPN}-1=(2.1 \pm 2.3) \times 10^{-5}$ \cite{Bertotti:2003rm}.
Since $\gamma_{\rm PPN}-1$ is negative in the current theory, 
we adopt the limit $|\gamma_{\rm PPN}-1| \leq 0.2 \times 10^{-5}$.
This corresponds to the bound $|\beta \varphi_0| \le 1 \times 10^{-3}$, 
so today's field value $\phi_0$ is constrained to be
\be
|\phi_0| \le 1.4 \times 10^{-3}\Mpl |\beta|^{-1}\,.
\label{phi0up}
\ee
For $\beta=-4.4$, we have $|\phi_0| \leq 3.2 \times 10^{-4} \Mpl$. 
In the following, we will study how this constraint translates to 
the upper limit of $|\phi|$ at the onset of radiation era.
 
After the reheating period ends, we can neglect the contribution 
of the inflaton field $\chi$ in Eq.~(\ref{meff}).
Moreover, provided that the energy density of $\phi$ is negligible 
relative to that of the background and that $|\phi| \ll \Mpl$, 
the effective mass squared (\ref{meff}) is 
approximated as $m_{\rm eff}^2 \simeq 3 \beta (2H^2 +\dot{H})$.
Then, the scalar-field equation (\ref{phieq}) approximately reduces to 
\be
\phi''+\frac{3}{2} \left( 1 -w_{\rm eff} \right) \phi'+\frac{3}{2} 
 \left( 1 - 3w_{\rm eff} \right) \beta \phi \simeq 0\,,
\ee
where a prime represents the derivative with respect to $N=\ln a$, 
and $w_{\rm eff}=-1-2H'/(3H)$ is the effective equation of state. 
If $w_{\rm eff}={\rm constant}$, there is the following 
growing-mode solution 
\be
\phi \propto a^{\lambda}\,,
\ee
where 
\be
\lambda=\frac{3}{4} \left( 1-w_{\rm eff} \right) 
\left[ \sqrt{1-\frac{8\beta (1-3w_{\rm eff})}
{3(1-w_{\rm eff})^2}} -1\right].
\ee
During the radiation era ($w_{\rm eff}=1/3$), we have $\lambda=0$ 
and hence $\phi={\rm constant}$. This property is attributed to the fact that 
$R$ vanishes on the exact radiation-dominated background.
In the matter-dominated era ($w_{\rm eff}=0$), the negative mass squared 
of $\phi$ leads to the following tachyonic growth of the field
\be
\phi \propto a^{(3/4)\left( \sqrt{1-8\beta/3} 
-1 \right)}\,.
\label{phiap}
\ee
During the epoch of cosmological constant domination, the growth of 
$\phi$ is even stronger: $\phi \propto 
a^{(3/2)\left( \sqrt{1-8\beta/3} -1 \right)}$. 
However, the dominance of dark energy over nonrelativistic matter 
occurs only at low redshifts $z \lesssim 0.3$, so we can approximately 
use Eq.~(\ref{phiap}) for the evolution of $\phi$ from radiation-matter 
equality to today. Then, the field value at radiation-matter equality 
can be estimated as 
\be
\phi_{\rm eq}=\phi_0 (z_{\rm eq}+1)^{-(3/4)\left( \sqrt{1-8\beta/3} 
-1 \right)}\,,
\label{phiana}
\ee
where $z_{\rm eq}$ is determined by 
$z_{\rm eq}=\Omega_{m0}/\Omega_{r0}-1$, with 
$\Omega_{m0}$ and $\Omega_{r0}$ being today's density parameters of 
nonrelativistic matter and radiation respectively. 

\begin{figure}[h]
\begin{center}
\includegraphics[height=3.3in,width=3.4in]{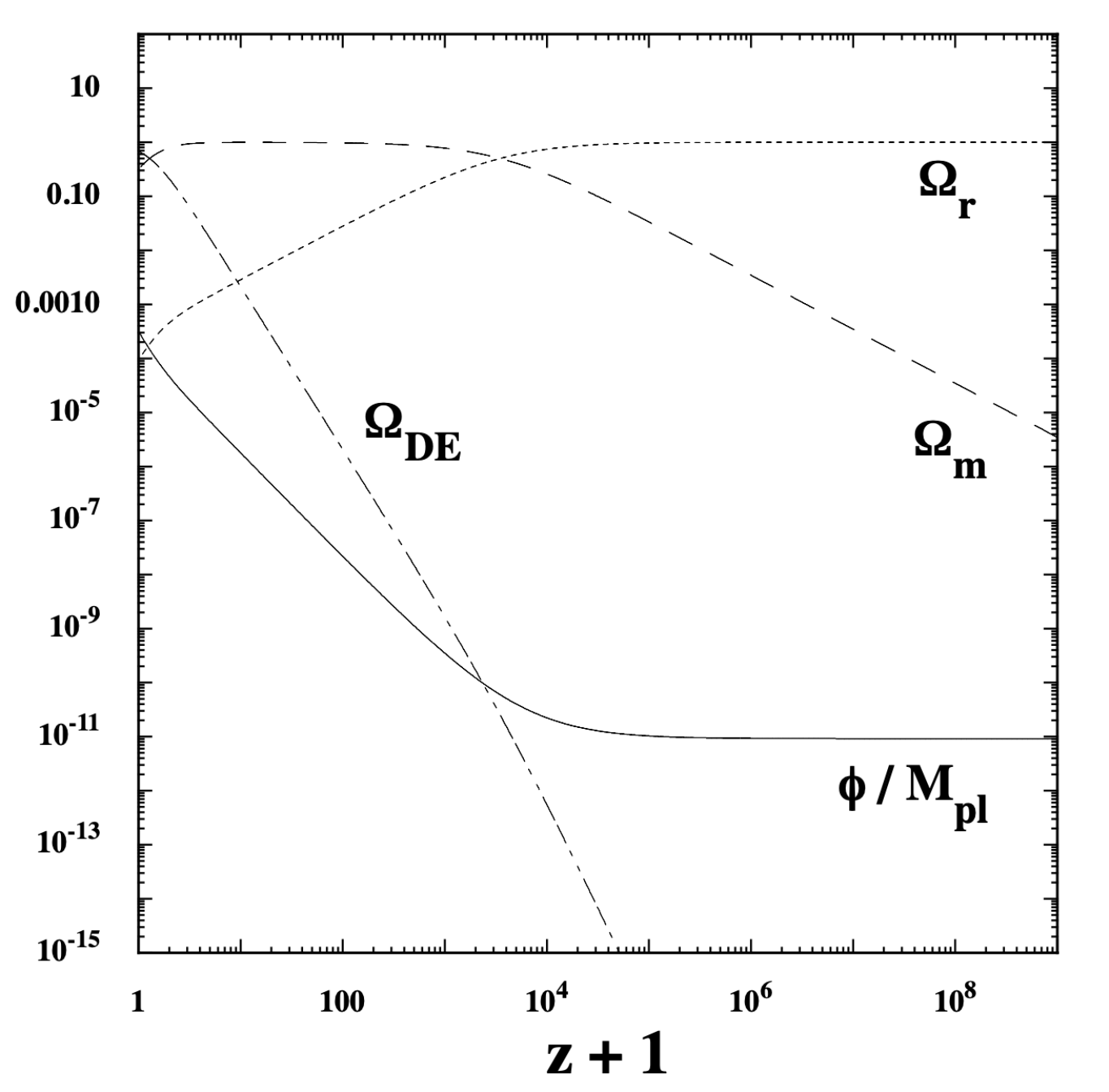}
\end{center}
\caption{\label{fig1}
Evolution of $\phi/M_{\rm pl}$ and $\Omega_r$, 
$\Omega_m$, and $\Omega_{\rm DE}$ versus $z+1~(=1/a)$  
for $\beta=-4.4$. 
The initial conditions are chosen to realize today's values 
$\phi_0/M_{\rm pl}=3.2 \times 10^{-4}$, 
$\Omega_{r0}=9.2 \times 10^{-5}$, and 
$\Omega_{{\rm DE}0}=0.68$.
}
\end{figure}

In Fig.~\ref{fig1}, we plot the evolution of 
$\phi/M_{\rm pl}$ and $\Omega_r=\rho_r/(3F H^2 \Mpl^2)$, 
$\Omega_m=\rho_m/(3F H^2 \Mpl^2)$, and 
$\Omega_{\rm DE}=\Lambda/(3F H^2 \Mpl^2)$ for $\beta=-4.4$ 
between the radiation era and today.
We choose the initial conditions around redshift $z=10^9$ 
to realize today's value $\phi_0=3.2 \times 10^{-4}\Mpl$, 
which corresponds to the upper limit 
consistent with local gravity constraints.
In this case the redshift at radiation-matter equality is 
$z_{\rm eq} \simeq 3470$, so the analytic estimation (\ref{phiana}) gives 
$\phi_{\rm eq}=4.8 \times 10^{-11}\Mpl$. 
This is fairly close to the numerical value 
$\phi_{\rm eq}=6.0 \times 10^{-11}\Mpl$, even though 
we ignored the epoch of late-time cosmic acceleration 
for the analytic estimation of $\phi_0$.

In Fig.~\ref{fig1}, we observe that $\phi$ slowly grows even at the 
late stage of radiation era. This reflects the fact that 
$R=6(2H^2+\dot{H})$ does not completely vanish due to 
the presence of nonrelativistic matter. 
Numerically, we find that $\phi_{\rm eq}$ is larger than the initial 
value $\phi_{\rm R}$ in the radiation era by one order of magnitude. 
This is consistent with the analytic estimation given in Ref.~\cite{Anson:2019ebp}. 
Multiplying the analytic value (\ref{phiana}) by a factor 0.2, it is possible to 
reproduce the numerical value of $\phi_{\rm R}$ approximately.
Then, we have the following analytic formula
\be
|\phi_{\rm R}| \simeq 0.2 |\phi_0| \left( \frac{\Omega_{m0}}
{\Omega_{r0}} \right)^{-(3/4)\left( \sqrt{1-8\beta/3} 
-1 \right)}\,.
\label{phiR}
\ee
For the model parameters used in the numerical simulation of Fig.~\ref{fig1}, i.e., 
$\phi_0=3.2 \times 10^{-4}\Mpl$, 
$\Omega_{m0}=0.32$, and $\Omega_{r0}=9.2 \times 10^{-5}$,
the analytic estimation (\ref{phiR}) gives 
$|\phi_{\rm R}|=9.6 \times 10^{-12}\Mpl$.
This is close to the numerical value $|\phi_{\rm R}|=9.2 \times 10^{-12}\Mpl$. 

Applying the formula (\ref{phiR}) to Eq.~(\ref{phi0up}), we obtain 
the following upper bound
\be
|\phi_{\rm R}| \le 2.8 \times 10^{-4} \Mpl |\beta|^{-1} 
\left( \frac{\Omega_{m0}}
{\Omega_{r0}} \right)^{-(3/4)\left( \sqrt{1-8\beta/3} 
-1 \right)}.
\label{phiR2}
\ee
For $\beta=-4.5$ and $\beta=-4.35$, the criterion (\ref{phiR2}) gives $|\phi_{\rm R}| \le 7.5 \times 10^{-12}\Mpl$ 
and $|\phi_{\rm R}| \le 1.1 \times 10^{-11}\Mpl$, respectively, where we used the same values of  $\Omega_{m0}$ and $\Omega_{r0}$ mentioned above.
These bounds are close to the numerically derived upper limits 
$|\phi_{\rm R}| \le 7.1 \times 10^{-12}\Mpl$ 
and $|\phi_{\rm R}| \le 1.1 \times 10^{-11}\Mpl$, respectively.
Thus, in the coupling range $-4.5 \le \beta \le -4.35$, 
the initial value of $|\phi_{\rm R}|$ at the onset of 
radiation era needs to be smaller than the order $10^{-11}\Mpl$ 
for the consistency with local gravity constraints.
 
\section{Inflationary epoch}
\label{infsec}

In this section, we study the evolution of $\phi$ during inflation 
in the presence of the coupling $(1/2)g^2 \phi^2 \chi^2$ 
besides the nonminimal coupling (\ref{Fnon}) with $\beta<0$. 
We consider two inflaton potentials: 
(i) $V(\chi)=m^2 \chi^2/2$ and
(ii) $\alpha$-attractor potential (\ref{Vchi}) with $\alpha=1$. 
The potential (i), which corresponds to the limit $\alpha \to \infty$ 
of Eq.~(\ref{Vchi}), leads to the scalar spectral index 
$n_s \simeq 1-2/N$ 
and the tensor-to-scalar ratio 
$r \simeq 8/N$, where $N$ is the 
number of e-foldings backward from the end of inflation to the epoch 
at which the perturbations relevant to observed 
Cosmic Microwave Background (CMB) temperature 
anisotropies crossed the Hubble radius \cite{Bassett:2005xm}. 
For $N=60$ we have $n_s \simeq 0.967$ and $r=0.133$, so the 
model is in tension with the Planck2018 bound 
of the tensor-to-scalar ratio $r<0.066$ (95\,\% CL) \cite{Planck:2018jri}.

Nevertheless, the potential (i) of chaotic inflation with the four-point interaction 
$g^2 \phi^2 \chi^2/2$ is a baseline model widely studied in the context of 
preheating after inflation \cite{Kofman:1994rk,Shtanov:1994ce,Kaiser:1995fb,Khlebnikov:1996zt,Khlebnikov:1996wr,Prokopec:1996rr,Kofman:1997yn,Bassett:2005xm}. 
We will accommodate this case in our analysis for the purpose of understanding the difference from the $\alpha$-attractor with $\alpha=1$.
In the model (ii) with $\alpha \lesssim {\cal O}(1)$, 
we have  $n_s \simeq 1-2/N$ and 
$r \simeq 12 \alpha/N^2$ \cite{Kallosh:2013yoa}, 
and hence $r=3.3 \times 10^{-3}$ for $\alpha=1$ and $N=60$. 
If $\alpha \lesssim 40$, the $\alpha$-attractor model is compatible 
with the Planck CMB bound $r<0.066$ and $n_s=0.9661 \pm 0.0040$ 
(68\,\% CL) \cite{Heisenberg:2018erb}.

In the current theory the field $\phi$ is present besides the inflaton $\chi$, 
so the existence of the former field can modify the prediction of inflationary observables like $n_s$ and $r$. 
To avoid this, we consider the case in which the contribution 
of $\phi$ to the background equations of motion 
is suppressed around the e-folding $N=60$ backward from the end of inflation. 
This amounts to the conditions $\beta^2 \phi^2/(2\Mpl^2) \ll 1$, 
$\dot{\phi}^2 \ll H^2 \Mpl^2$, and $g^2 \phi^2 \chi^2 \ll V$. 
Considering the coupling $\beta$ in the range $-4.5 \le \beta \le -4.35$ with $|\dot{\phi}|$ at most of order $|H\phi|$, the first two conditions 
can be satisfied for $|\phi| \lesssim 0.1 \Mpl$.  
Then, the field value $\phi_{\rm inf}$ about the $60$ e-folding 
backward from the end of inflation is in the range
\be
|\phi_{\rm inf}| \lesssim 0.1 \Mpl\,,\quad {\rm and} \quad 
|\phi_{\rm inf}| \lesssim 0.1 \frac{\sqrt{V}}{g|\chi|}\,.
\label{phiinf}
\ee
For the potential $V(\chi)=m^2 \chi^2/2$, the latter condition 
translates to $|\phi_{\rm inf}| \lesssim 0.1m/g$.

Under the conditions (\ref{phiinf}), the effective mass squared (\ref{meff}) 
during inflation ($|\dot{H}| \ll H^2$ and $3H^2 \Mpl^2 \simeq V$) 
approximately reduces to 
\be
m_{\rm eff}^2 \simeq g^2 \chi^2+6 \beta H^2 
\simeq g^2 \chi^2+\frac{2 \beta V}{\Mpl^2}\,.
\label{meffinf}
\ee
When $g=0$, we have 
$m_{\rm eff}^2<0$ and hence 
there is the tachyonic growth of $\phi$ \cite{Anson:2019ebp}.
On the exact de-Sitter background, the growing-mode 
solution to Eq.~(\ref{phieq}) for $g=0$ is given by 
\be
\phi \propto \exp \left[ \frac{3}{2} 
\left( \sqrt{1-\frac83 \beta}-1 \right) H t 
\right]\,.
\ee
During the time interval $t=10H^{-1}$, for instance, the field $\phi$ is 
amplified by a factor $5 \times 10^{16}$ for $\beta=-4.4$. 
This enhancement of $\phi$ destroys the inflationary period 
driven by the potential energy of $\chi$. 
Even if the initial field value $\phi_{\rm inf}$ is fine-tuned to be 
extremely close to 0, 
there exists a field perturbation $\delta \phi$ whose amplitude 
does not vanish due to the uncertainty principle.  
After the Hubble radius crossing, the perturbation $\delta \phi$ is 
amplified in a manner analogous to the homogeneous field $\phi$ 
discussed above \cite{Anson:2019ebp}.
Hence the existence of the nonminimal coupling (\ref{Fnon}) 
with $\beta=-{\cal O}(1)$ 
violates the successful inflationary prediction of primordial density perturbations sourced by the perturbations of $\chi$.

The nonvanishing coupling $g$ gives rise to a positive 
contribution to the mass squared (\ref{meffinf}). 
If $g$ is in the range
\be
g > \frac{\sqrt{2|\beta|V}}{\Mpl |\chi|}\,,
\label{gcon}
\ee
we have $m_{\rm eff}^2 > 0$ and hence the exponential growth 
of $|\phi|$ can be avoided.  

Let us first consider the effective mass in the range 
$0 < m_{\rm eff}^2 \leq 9H^2/4$.
Neglecting the variation of $m_{\rm eff}^2$ during inflation, 
the growing-mode solution to Eq.~(\ref{phieq}) is given by 
\be
\phi \propto \exp \left[ -\frac{3}{2} 
\left( 1-\sqrt{1-\frac{4m_{\rm eff}^2}{9H^2}} \right) 
Ht \right]\,.
\label{phide}
\ee
When $m_{\rm eff}^2=0$ we have $\phi={\rm constant}$, 
while, for increasing $m_{\rm eff}^2$, 
the decreasing rate of $\phi$ gets larger.

For the effective mass satisfying $m_{\rm eff}^2>9H^2/4$, 
the coupling $g$ is in the range 
\be
g>\frac{\sqrt{(3-8\beta)V}}{2\Mpl |\chi|}\,,
\label{gcon2}
\ee
whose lower limit is larger than that in Eq.~(\ref{gcon}). 
In this mass range, the field $\phi$ evolves as
\be
\phi \propto \exp \left( -\frac{3}{2}Ht \right) 
\cos \left( \sqrt{1-\frac{9H^2}{4m_{\rm eff}^2}}\,
m_{\rm eff} t \right)\,.
\ee
Then, the amplitude of $\phi$ decreases as
$|\phi| \propto e^{-3Ht/2} \propto a^{-3/2}$ 
with the oscillation induced by 
the effective mass $m_{\rm eff}$ \cite{Jedamzik:1999um,Ivanov:1999hz,Liddle:1999hq,Tsujikawa:2002nf}, 
where we used the approximate solution $a \propto e^{Ht}$ 
during inflation. 
On using the number of e-foldings $N_{\rm inf}$ relevant to the 
observation of CMB temperature anisotropies, the amplitude of 
$\phi$ at the end of inflation (denoted as $|\phi_I|$) 
can be estimated as
\be
|\phi_I|= |\phi_{\rm inf}| e^{-3N_{\rm inf}/2}\,.
\label{phiI}
\ee
For $N_{\rm inf}=60$ and $\phi_{\rm inf}=0.1 \Mpl$, 
the estimation (\ref{phiI}) gives $|\phi_I|=8 \times 10^{-41}\Mpl$, 
so there is the strong suppression of $|\phi|$ during inflation.
Notice that, for $m_{\rm eff}^2>9H^2/4$, the decreasing rate of 
$|\phi|$ does not depend on $g$.

\subsection{Quadratic potential}
\label{Quasec}

For the inflaton potential $V(\chi)=m^2 \chi^2/2$, the inequalities
(\ref{gcon}) and (\ref{gcon2}) translate to 
$g > \sqrt{|\beta|}m/\Mpl$ and $g>\sqrt{3/8-\beta}\,m/\Mpl$, 
respectively. We consider the mass $m=6 \times 10^{-6}\Mpl$ 
constrained from the Planck normalization of CMB 
temperature anisotropies \cite{Planck:2018jri}. 
For $\beta=-4.4$, the conditions 
(\ref{gcon}) and (\ref{gcon2}) correspond to 
$g>1.26 \times 10^{-5}$ and $g>1.31 \times 10^{-5}$, respectively.

\vspace{0.5cm}
\begin{figure}[h]
\begin{center}
\includegraphics[height=3.1in,width=3.4in]{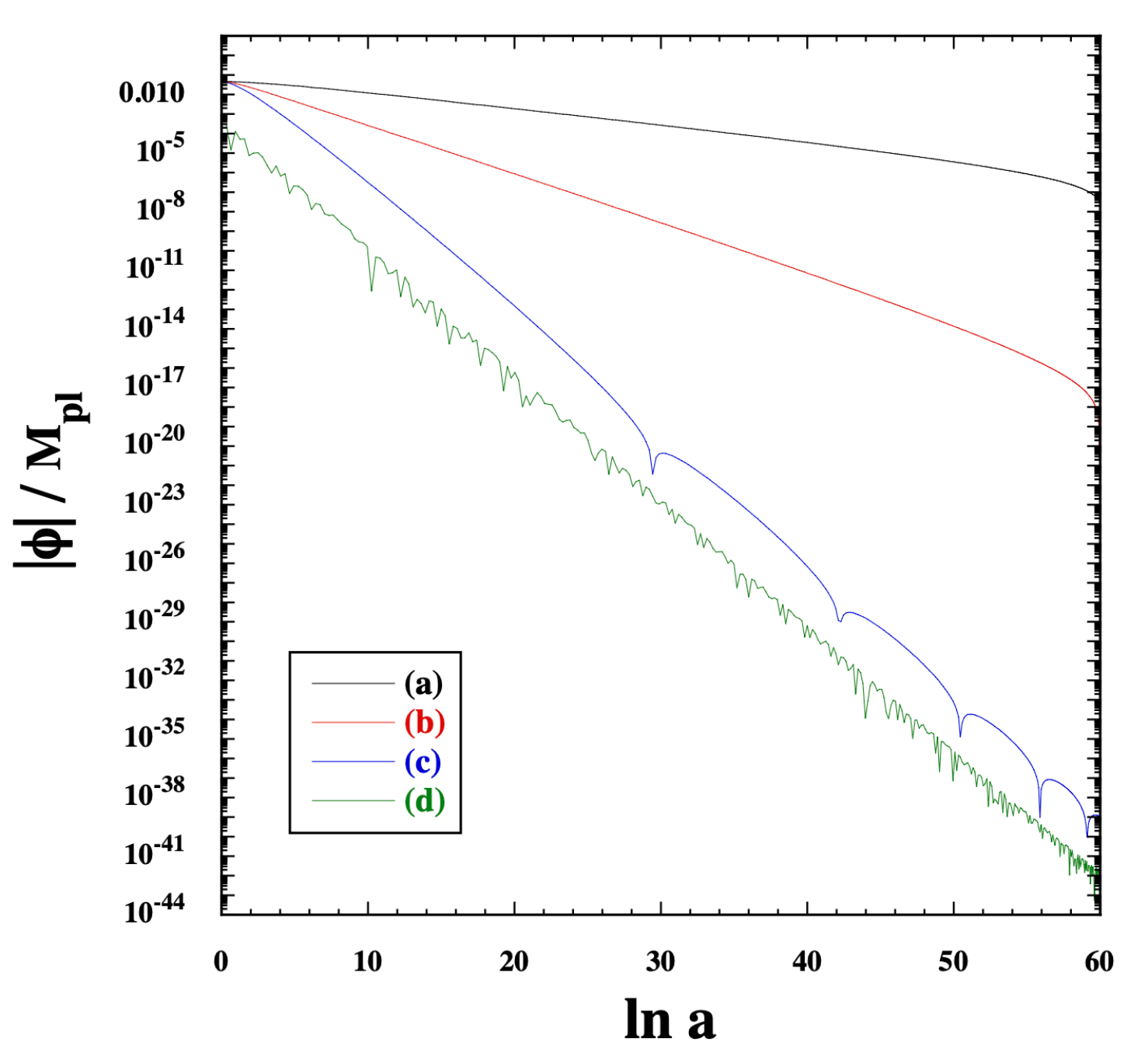}
\end{center}
\caption{\label{fig2}
Evolution of $|\phi|/\Mpl$ versus $\ln a$ during inflation for 
the potential $V(\chi)=m^2 \chi^2/2$ with $\beta=-4.4$ 
and $m=6.0 \times 10^{-6}\Mpl$. 
The initial field values of scalar fields are chosen to be
$\chi=15.38\Mpl$, 
with $\phi$ corresponding to the maximum of Eq.~(\ref{phiinf}). 
In this case, the number of e-foldings during inflation 
is $N=60$.
Each line corresponds to (a) $g=1.27 \times 10^{-5}$, 
(b) $g=1.29 \times 10^{-5}$, (c) $g=1.31 \times 10^{-5}$, 
and (d) $g=1.0 \times 10^{-3}$.
}
\end{figure}

In Fig.~\ref{fig2}, we plot the evolution of $|\phi|/\Mpl$ for four different coupling constants $g$. We choose the initial field 
value $\phi_{\rm inf}$ as a maximum satisfying the two conditions given in Eq.~(\ref{phiinf}).
In case (a) the coupling $g=1.27 \times 10^{-5}$ is 
slightly larger 
than the value $g=1.26 \times 10^{-5}$ corresponding to $m_{\rm eff}^2=0$, 
so the field $\phi$ mildly decreases according to Eq.~(\ref{phide}). 
As $g$ increases in the range $1.26 \times 10^{-5}<g<1.31 \times 10^{-5}$, 
the decreasing rate of $\phi$ tends to be larger as we observe
in cases (b) and (c) of Fig.~\ref{fig2}. 
For $g>1.31 \times 10^{-5}$, the amplitude of $\phi$ 
decreases in proportion to $a^{-3/2}$ with oscillations. 
Thus, provided the condition (\ref{gcon2}), i.e., 
\be
g>\sqrt{\frac{3}{8}-\beta}\,\frac{m}{\Mpl},
\label{gconm}
\ee
is satisfied, the field value at the end of inflation is 
suppressed 
to be in the range $|\phi_I| \lesssim 10^{-40}\Mpl$. 
For such couplings, unless $|\phi|$ is amplified by a factor 
more than $10^{29}$ during reheating, the bound 
$|\phi_{\rm R}| \lesssim 10^{-11}\Mpl$ can be satisfied 
at the onset of radiation era.
For the coupling $g$ in the range 
$\sqrt{2|\beta|V}/(\Mpl |\chi|)<g<\sqrt{(3-8\beta)V}/(2\Mpl |\chi|)$, 
the suppression of $\phi$ during inflation is not necessarily significant, 
so the growth of $\phi$ during reheating matters to 
satisfy the bound $|\phi_{\rm R}| \lesssim 10^{-11}\Mpl$.

\subsection{$\alpha$-attractor with $\alpha=1$}
\label{alinfs}

Let us consider the $\alpha$-attractor potential (\ref{Vchi}) with $\alpha=1$. 
In this case, the right hand-sides of Eqs.~(\ref{gcon}) and (\ref{gcon2}) 
depend on the value of $\chi$. Under the slow-roll approximation,
the number of e-foldings backward from the end of inflation 
(inflaton value $\chi_f$) can be computed by 
$N=\Mpl^{-2} \int_{\chi_f}^{\chi} V/V_{,\chi}\,{\rm d}\chi$. 
In the present case, we have $N=(3/4)[1/y-1/y_f+\ln (y/y_f)]$, 
where $y=e^{-\sqrt{2/3}\chi/\Mpl}$ and 
$y_f=e^{-\sqrt{2/3}\chi_f/\Mpl}$. 
The field value at the end of inflation is determined by 
the condition $\epsilon_V=(\Mpl^2/2)(V_{,\chi}/V)^2=1$, 
and hence $\chi_f=0.940 \Mpl$ and $y_f=0.464$. 
For $N=60$, we have $y=1.165 \times 10^{-2}$ and 
$\chi=5.453 \Mpl$, in which case the Planck normalization 
of primordial curvature perturbations gives 
the constraint $m \simeq 1.1 \times 10^{-5}\Mpl$.

Since $\chi \gtrsim {\cal O}(\Mpl)$ during the inflationary period, 
the potential (\ref{Vchi}) is nearly constant in this regime and 
hence the term $g^2 \chi^2$ in Eq.~(\ref{meffinf}) decreases faster 
than $2|\beta| V/\Mpl^2$. 
Unlike the quadratic potential, there are some ranges of 
the coupling $g$ for which $g^2 \chi^2$ is initially larger 
than $2|\beta|V/\Mpl^2$, but the latter dominates 
over the former during inflation. 
On using the value $\chi=5.453 \Mpl$ at $N=60$ with 
$\beta=-4.4$ and $m=1.1 \times 10^{-5}\Mpl$, 
the condition (\ref{gcon}) for the realization of 
positive $m_{\rm eff}^2$ translates to  $g>5.12 \times 10^{-6}$.

\begin{figure}[h]
\begin{center}
\includegraphics[height=3.1in,width=3.4in]{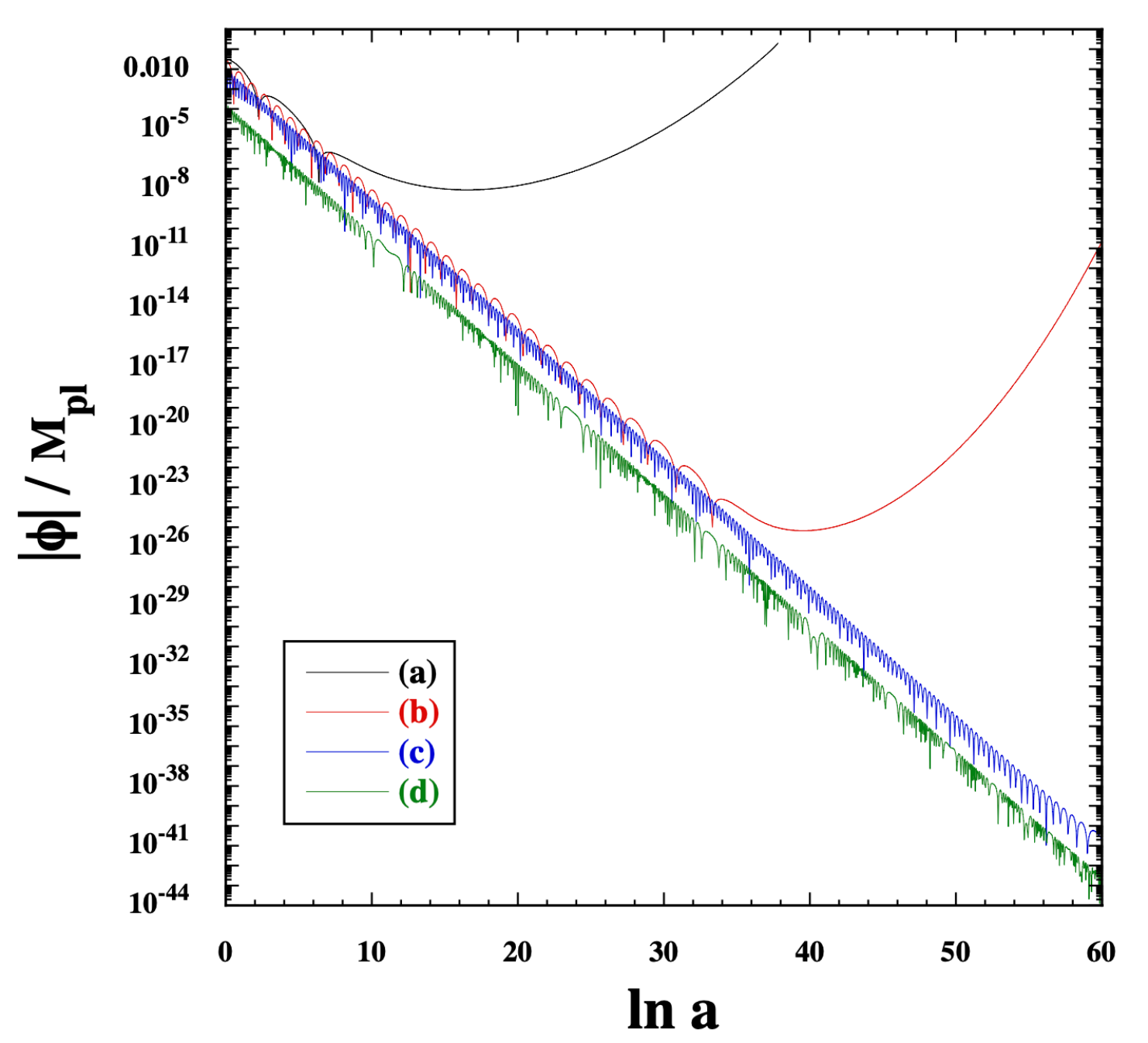}
\end{center}
\caption{\label{fig3}
Evolution of $|\phi|/\Mpl$ versus $\ln a$ during inflation for 
the $\alpha$-attractor potential (\ref{Vchi}) with $\alpha=1$, 
$\beta=-4.4$, and $m=1.1 \times 10^{-5}\Mpl$. 
The initial field values of scalar fields are chosen to be $\chi=5.418\Mpl$, 
with $\phi$ corresponding to the maximum of Eq.~(\ref{phiinf}).
Each case corresponds to (a) $g=5.5 \times 10^{-6}$, 
(b) $g=6.5 \times 10^{-6}$, (c) $g=1.68 \times 10^{-5}$, 
and (d) $g=1.0 \times 10^{-3}$.
}
\end{figure}

In case (a) of Fig.~\ref{fig3}, we plot the evolution of $|\phi|/\Mpl$ 
during inflation for $g=5.5 \times 10^{-6}$. 
In this case the field initially decreases because $m_{\rm eff}^2>0$, 
but it starts to grow at some point 
because $g^2 \chi^2$ drops below $2|\beta|V/\Mpl^2$. 
This enhancement of $\phi$ largely modifies the dynamics 
of inflation in such a way that the total number of e-foldings 
does not reach even 40. As we observe in case (b) of Fig.~\ref{fig3}, 
the growth of $\phi$ also occurs for the coupling 
$g=6.5 \times 10^{-6}$. In this case, the field value $\phi$ 
at the end of inflation is of order $10^{-11}\Mpl$, so the 
dynamics of inflation driven by the field $\chi$ is hardly modified. 
In Sec.~\ref{reheatsec}, we will study whether or not the existence of a 
subsequent reheating stage leads to additional 
growth of $\phi$ exceeding the bound $|\phi_{\rm R}|<10^{-11}\Mpl$.

If we demand that the condition (\ref{gcon}), i.e., $m_{\rm eff}^2>0$,  
holds by the end of inflation (at which $\chi_f=0.940 \Mpl$), 
then the coupling $g$ is constrained to be 
\be
g>0.698 \sqrt{|\beta|} \frac{m}{\Mpl}\,.
\label{gcona}
\ee
Substituting the values $\beta=-4.4$ and 
$m=1.1 \times 10^{-5}\Mpl$ into Eq.~(\ref{gcona}), we have
$g>1.61 \times 10^{-5}$. Similarly, the condition (\ref{gcon2}) 
translates to $g>1.68 \times 10^{-5}$. 
The evolution of $|\phi|/M_{\rm pl}$ for $g=1.68 \times 10^{-5}$ 
is plotted as case (c) in Fig.~\ref{fig3}. 
We observe that the field exhibits exponential 
decrease $|\phi| \propto a^{-3/2}$ by the end of inflation. 
Numerically, we find that, even for the marginal coupling 
$g=1.61 \times 10^{-5}$, $\phi$ decreases in a similar manner to case (c). 
Hence the condition (\ref{gcona}) is practically sufficient to ensure 
the exponential suppression of $\phi$ during the whole stage of inflation. 
As we see in case (d), the decreasing rate of $\phi$ for $g=1.0 \times 10^{-3}$ 
is practically the same as in case (c). 
The only difference between cases (c) and (d) is the choice of 
initial conditions, where we have selected the maximum 
value of $|\phi_{\rm inf}|$ satisfying the conditions (\ref{phiinf}).
For larger $g$, $|\phi_{\rm inf}|$ tends to be smaller.

\section{Reheating epoch}
\label{reheatsec}

After the end of inflation, the Universe enters a reheating stage 
in which the inflaton field $\chi$ oscillates around the 
potential minimum. 
The field value $\chi_I$ at the onset of reheating is determined by 
the condition $\epsilon_V=(\Mpl^2/2)(V_{,\chi}/V)^2=1$, so that 
$\chi_I=1.414 \Mpl$ for $V(\chi)=m^2 \chi^2/2$ 
and $\chi_I=0.940 \Mpl$ for the $\alpha$-attractor potential 
(\ref{Vchi}) with $\alpha=1$. 
Around $\chi=0$, the potential (\ref{Vchi}) 
approximately reduces to the quadratic one: 
$V(\chi) \simeq m^2 \chi^2/2$.

In the presence of the coupling $g^2 \phi^2 \chi^2/2$, 
the background field $\phi$ and its perturbations can be 
resonantly amplified by broad parametric resonance 
during reheating for the coupling 
$g$ in the range $g^2 \chi^2 \gg m^2$ \cite{Kofman:1994rk,Kofman:1997yn}.
The resonant growth of the variance of $\phi$ affects the evolution of the homogeneous mode of $\chi$ through the backreaction of created 
particles \cite{Khlebnikov:1996wr,Kofman:1997yn,Tsujikawa:1999jh}. 
To incorporate this effect, we expand the inhomogeneous field 
$\delta \phi(t,{\bm x})$ 
in terms of the Fourier series, as
\be
\delta \phi(t,{\bm x})=\frac{1}{(2\pi)^3} 
\int {\rm d}^3 k\, 
\delta \phi_k (t) 
e^{i {\bm k} \cdot {\bm x}}\,,
\ee
where ${\bm k}$ is a comoving wavenumber with $k=|{\bm k}|$. 
Under a Hartree approximation, 
the zero-momentum 
mode of $\chi$ is the only nonvanishing component 
of inflaton \cite{Khlebnikov:1996wr}. 
Using this approximation with $V(\chi) \simeq m^2 \chi^2/2$ 
in the reheating stage, Eq.~(\ref{chieq}) is modified to 
\be
\ddot{\chi}+\left( 3H +\Gamma \right) \dot{\chi}
+\left[ m^2+g^2 \left(\phi^2+ 
\langle \delta \phi^2 \rangle \right) \right] 
\chi=0\,,
\label{chiback}
\ee
where $\phi$ is the homogeneous 
value of the field, and 
$\langle \delta \phi^2 \rangle$ is the vacuum expectation 
value given by 
\be
\langle \delta \phi^2 \rangle=\int \frac{{\rm d}k}{k}{\cal P}_{\delta\phi_k}\,,
\label{phisq}
\ee
with the power spectrum
\be
{\cal P}_{\delta\phi_k}=\frac{k^3}{2\pi^2} |\delta \phi_k|^2\,.
\label{Pchik}
\ee

The Fourier modes $\delta \phi_k (t)$ obey the differential equation
\be
\ddot{\delta \phi}_k+\left( 3H+\frac{\omega_{,\phi} \dot{\phi}}
{\omega} \right) \dot{\delta\phi}_k
+\left( \frac{k^2}{a^2}+M_{\rm eff}^2 \right) 
\delta \phi_k=0\,,
\label{phikeq}
\ee
where
\ba
M_{\rm eff}^2 &=&
\frac{1}{\omega} \biggl[ g^2 \chi^2
-3F_{,\phi \phi} \Mpl^2 (2H^2+\dot{H}) \nonumber \\
& &
~~~~+( \ddot{\phi}+3H \dot{\phi} ) \omega_{,\phi}
+\frac{\dot{\phi}^2}{2} \omega_{,\phi \phi} \biggr].
\label{Meff}
\ea
To the right hand-sides of Eqs.~(\ref{back1}) and 
(\ref{back2}), we take into account the perturbed density 
$\langle \rho_{\delta \phi} \rangle$ and the pressure 
$\langle P_{\delta \phi} \rangle$ of produced $\phi$ 
particles, respectively. 
Assuming that the conditions 
$m \gg H$ and 
$\langle \delta \phi^2 \rangle 
\ll \Mpl^2$ hold during reheating, 
we have
\ba
\hspace{-0.5cm}
\langle \rho_{\delta \phi} \rangle
&=&\frac{\omega(\phi)}{2} 
\left[ \langle \dot{\delta \phi}^2 \rangle+\frac{1}{a^2}
\langle \partial \delta \phi \rangle^2  \right]
+\frac{1}{2} g^2 \chi^2 \langle \delta \phi^2 \rangle,\\
\hspace{-0.5cm}
\langle P_{\delta \phi} \rangle
&=& \left[
\frac{\omega(\phi)}{2}
-\beta F(\phi) \right]
\left[ \langle \dot{\delta \phi}^2 \rangle+\frac{1}{a^2}
\langle \partial \delta \phi \rangle^2  \right] \nonumber \\
& &
-\frac{1}{2} g^2 \chi^2 \langle \delta \phi^2 \rangle \,,
\ea
where
\ba
\langle \dot{\delta \phi}^2 \rangle 
&=& \int {\rm d}k \frac{k^2}{2\pi^2}
|\dot{\delta \phi}_k |^2 \,,\label{vari1}\\
\langle \partial \delta \phi \rangle^2
&=& \int {\rm d}k \frac{k^4}{2\pi^2} \left| \delta \phi_k \right|^2\,.\label{vari2}
\ea

Defining the rescaled perturbed field 
\be
\delta \varphi_k=a^{3/2} \omega^{1/2} \delta \phi_k\,,
\ee
Eq.~(\ref{phikeq}) can be expressed in the form 
\be
\ddot{\delta \varphi}_k+\Omega_k^2\,\delta \varphi_k=0\,,
\ee
where
\ba
\Omega_k^2 &=& \frac{k^2}{a^2}-\frac{9}{4}H^2-\frac{3}{2} \dot{H}
+\frac{1}{\omega} \biggl[ g^2 \chi^2
-3F_{,\phi \phi} \Mpl^2 (2H^2+\dot{H}) \nonumber \\
& &
+\frac{1}{2}( \ddot{\phi}+3H \dot{\phi} ) \omega_{,\phi}
+\frac{\dot{\phi}^2 \omega_{,\phi}^2}{4\omega} \biggr]\,.
\label{Omek}
\ea
The typical wavenumber relevant to the parametric excitation of 
$\phi$ particles is $k/a_I \sim m$, where $a_I$ is the scale factor 
at the onset of reheating.
The perturbations with $k/a_I \sim m$ are deep inside the Hubble 
radius during inflation ($k/a \gg H$), so the dominant 
contribution to $\Omega_k^2$ is the term $k^2/a^2$. 
For such modes, we choose a positive-frequency 
solution in the Bunch-Davies vacuum state as 
\be
\delta \varphi_k=\frac{1}{\sqrt{2\Omega_k}} 
e^{-i \int \Omega_k {\rm d}t}\,,
\label{phiini}
\ee
which corresponds to an initial condition of $\delta \varphi_k$
at the onset of reheating. 
For the modes $k/a_I \sim m$, the frequency $\Omega_k$ can be 
approximated as $\Omega_k \simeq k/a$ during inflation 
except for the last short period in which $k^2/a^2$ 
drops below $M_{\rm eff}^2$.
Then, the amplitude of $\delta \phi_k$ around the beginning 
of reheating is estimated as
\be
|\delta \phi_k(t_I)| \simeq \frac{1}{a_I\sqrt{\omega}
\sqrt{2k}}\,.
\ee
The square root of the power spectrum (\ref{Pchik}) for the mode $k/a_I \simeq m$ is given by  
\be
\sqrt{{\cal P}_{\delta \phi_k}(t_I)} \simeq 
\frac{1}{\sqrt{\omega}} \frac{k}{2\pi a_I} \simeq
\frac{m}{2\pi}\,,
\ee
where we used the approximation $\omega \simeq 1$ in the second 
equality. The perturbation $\delta \phi_k$ excited by 
parametric resonance has a typical initial amplitude 
$m/(2\pi)$. For the quadratic potential $V=m^2 \chi^2/2$ 
and the $\alpha$-attractor potential (\ref{Vchi}) with 
$\alpha=1$, we have $\sqrt{{\cal P}_{\delta \phi_k}(t_I)} \approx 10^{-6}\Mpl$.

Let us also estimate the power spectrum of larger-scale modes of 
$\delta \phi_k$ that exit the Hubble radius during inflation. 
We are interested in the range of coupling $g$ where parametric 
resonance occurs during preheating, in which case 
$M_{\rm eff}^2>9H^2/4$ during inflation.
After the Hubble radius crossing during inflation ($k<aH$), 
$\Omega_k^2$ is of order $M_{\rm eff}^2 \simeq g^2 \chi^2/\omega$.
Around the onset of reheating, the amplitude of $|\delta \phi_k|$ 
can be estimated as 
\be
|\delta \phi_k|(t_I) \simeq \frac{1}{a_I^{3/2} 
\sqrt{\omega} \sqrt{2M_{\rm eff}}}\,.
\ee
This corresponds to the power spectrum 
\be
{\cal P}_{\delta \phi_k}(t_I) \simeq 
\frac{1}{4\pi^2 \omega M_{\rm eff}} 
\left( \frac{k}{a_I} \right)^3 \simeq 
\frac{1}{4\pi^2}\frac{H_{\rm inf}^3}{M_{\rm eff}}
e^{-3N_{\rm inf}}\,,
\label{Pphik}
\ee
where $H_{\rm inf}$ is the value of $H$ 
at the Hubble radius crossing. 
In the second equality of Eq.~(\ref{Pphik}), we have substituted 
$k=a_{\rm inf}H_{\rm inf}$ and used the number of 
e-foldings $N_{\rm inf}=\ln (a_I/a_{\rm inf})$. 
For the perturbations relevant to the observed CMB temperature 
anisotropies, we have $N_{\rm inf}=55 \sim 60$. 
Taking the value $N_{\rm inf}=60$, the amplitude of perturbations 
$\delta \phi_k$ at the beginning of reheating is of order
\be
\sqrt{{\cal P}_{\delta \phi_k}(t_I)} \simeq 10^{-40}
\frac{H_{\rm inf}}{\sqrt{r_M}}\,,
\ee
where $r_M \equiv 
M_{\rm eff}/H_{\rm inf}$. 
For the inflationary scale $H_{\rm inf} \simeq 10^{-4}\Mpl$, 
we have $\sqrt{{\cal P}_{\delta \phi_k}(t_I)} \simeq 10^{-44}\Mpl/\sqrt{r_M}$. 
This suppression of the large-scale modes of $\delta \phi_k$ is 
analogous to what happens for the homogeneous field $\phi$, 
see Eq.~(\ref{phiI}). The positive mass squared $M_{\rm eff}^2$ 
greater than the order $H_{\rm inf}^2$ leads to the exponential 
decrease of $\delta \phi_k$ for the perturbations that exit the 
Hubble radius long before the end of inflation. 
{}From the above discussion, the large-scale perturbations 
$\delta \phi_k$ with wavenumbers $k$ in the range $k/a_I \ll m$ 
hardly contribute to the vacuum expectation value 
$\langle \delta \phi^2 \rangle$. 

Provided the background field $\phi$ is suppressed 
during inflation, its contributions to Eqs.~(\ref{chieq}) 
and (\ref{back1})-(\ref{back2}) can be ignored 
in the early stage of reheating. 
The initial reheating period in which $\chi$ oscillates 
coherently can be identified by a temporal matter era 
where the scale factor evolves as $a \propto t^{2/3}$ 
with $H=2/(3t)$. For the approximate potential 
$V \simeq m^2 \chi^2/2$ around $\chi=0$, 
the background Eqs.~(\ref{chieq}) and (\ref{back1}) 
reduce to $\ddot{\chi}+3H \dot{\chi}+m^2 \chi \simeq 0$ 
and $4 \Mpl^2/(3t^2) \simeq \dot{\chi}^2/2+m^2 \chi^2/2$, 
respectively. 
The solution consistent with the virial relation $\langle \dot{\chi}^2/2 
\rangle=\langle m^2 \chi^2/2 \rangle$ averaged over oscillations 
is given by 
\be
\chi=\chi_I \frac{t_I}{t}\sin (mt)\,,\quad
{\rm with} \quad \chi_I \equiv  
\sqrt{\frac83} \frac{\Mpl}{mt_I}\,,
\ee
where $\chi_I$ is the amplitude at the onset of 
reheating ($t=t_I$). 
The inflaton field oscillates with the amplitude decreasing 
in proportion to $1/t$. 

Ignoring the contributions of time derivatives of $\phi$ 
and terms $-9H^2/4-3\dot{H}/2$ in Eq.~(\ref{Omek}) 
during the coherent oscillation of $\chi$, the perturbation 
$\delta \varphi_k$ obeys the Mathieu equation
\be
\frac{{\rm d}^2\delta\varphi_k}{{\rm d} z^2}+\left[ 
A_k-2q \cos(2z) \right] \delta \varphi_k=0\,,
\label{varphieq}
\ee
where 
\ba
A_k &=& 
\left( \frac{k}{ma} \right)^2+2q+\frac{2\beta}{3z^2}\,,\label{Ak}\\
q &=& q_I \left( \frac{t_I}{t} \right)^2\,,\qquad 
q_I=\frac{g^2 \chi_I^2}{4m^2}\,,\\
z &=& mt\,.
\ea
As we see in Eq.~(\ref{phieq}), the homogeneous mode 
$\varphi =a^{3/2} \phi$ satisfies the same form of 
equation as (\ref{varphieq}) with the limit $k/(ma) \to 0$ 
in Eq.~(\ref{Ak}). If the parameter $q_I$ is in the range  
$q_I \gg 1$, it is known that there is an epoch of 
the preheating period in which 
$\delta \phi_k$ and $\phi$ are amplified by 
broad parametric resonance \cite{Kofman:1994rk,Kofman:1997yn}.
Since $\chi_I$ is of order $\Mpl$ for the inflaton potentials 
discussed in Sec.~\ref{infsec}, preheating can occur 
in the coupling range
\be
g \gg \frac{m}{\Mpl}\,.
\label{gpre}
\ee
For the exact quadratic potential $V=m^2 \chi^2/2$, the mass $m$ is constrained to be $m=6.0 \times 10^{-6}\Mpl$ from the Planck normalization, 
so the condition (\ref{gpre}) translates to $g \gg 6.0 \times 10^{-6}$.
Since the parameter $q$ decreases as $q \propto 1/t^2$ 
due to cosmic expansion, the growth of the variance $\langle \delta \phi^2 \rangle$ actually occurs for  $g \gtrsim 10^{-4}$ \cite{Kofman:1997yn,Khlebnikov:1996wr,Khlebnikov:1996zt}.
In the $\alpha$-attractor model with $\alpha=1$, we have 
$m=1.1 \times 10^{-5}\Mpl$ from the Planck normalization, 
so the bound (\ref{gpre}) 
corresponds to $g \gg 1.1\times 10^{-5}$. 
In this case, the field value $\chi_I$ at the onset of reheating 
is smaller than that for the exact quadratic potential, so the larger 
coupling $g$ is required for the realization of broad 
parametric resonance.

For the coupling $g$ in the range (\ref{gpre}), the homogeneous field $\phi$ 
is exponentially suppressed ($|\phi| \propto a^{-3/2}$) during 
inflation. Since we are considering the coupling constant around 
$\beta \sim -4$, the term $2\beta/(3z^2)$ in Eq.~(\ref{Ak}) 
is less than the order 1 
for $z \gtrsim 1$. 
Provided that $g \gg m/\Mpl$, the condition $2q \gg 2|\beta|/(3z^2)$ is satisfied during preheating, so the nonminimal coupling 
$\beta$ hardly affects the dynamics of $\delta \phi_k$. 
If $g \sim m/\Mpl$, the parameter $2q$ is of the same 
order as $2|\beta|/(3z^2)$  and hence the coupling 
$\beta$ cannot be ignored. In such cases, however, the 
absence of parametric resonance does not lead to 
the growth of the field variance $\langle \delta \phi^2 \rangle$.
For the range of $g$ where the preheating does not occur,  
we do not take the backreaction into account.

For wavenumbers in the range $k/a_I \gg m$, the term 
$(k/ma_I)^2$ is much larger than 1.  
Although $k/(ma)$ decreases during reheating, the parametric 
excitation of $\delta \phi_k$ is not significant for  
$k/a_I$ very much larger than $m$.
We recall that the perturbations $\delta \phi_k$ with $k/a_I \ll m$ as well as the homogeneous 
mode $\phi$ are subject to the exponential suppression during inflation. 
Then, the main contribution to $\langle \delta \phi^2 \rangle$ 
at the beginning of preheating comes from the modes distributed 
around $k/a_I \sim m$. 
We compute the variances (\ref{phisq}) and (\ref{vari1})-(\ref{vari2})  
for the wavenumbers up to $k/a_I \lesssim 10^2 m$ 
by using the initial condition (\ref{phiini}) at the onset of reheating. 

\subsection{Quadratic potential}

Let us first study the reheating dynamics for the inflaton 
potential $V(\chi)=m^2 \chi^2/2$. 
As we showed in Sec.~\ref{Quasec}, the amplitude of $\phi$ 
decreases as $|\phi| \propto a^{-3/2}$ during inflation 
for $g>1.31 \times 10^{-5}$. 
Parametric resonance occurs for $g \gtrsim 10^{-4}$, 
in which regime the field value $\phi_I$ 
at the onset of reheating is determined by Eq.~(\ref{phiI}), 
where $\phi_{\rm inf}$ is limited as Eq.~(\ref{phiinf}). 
When $g \gtrsim 10^{-4}$, the second condition of (\ref{phiinf}) 
gives the upper bounds of $\phi_{\rm inf}$ and $\phi_I$.
For larger $g$, the maximum values of $\phi_{\rm inf}$ 
and $\phi_I$ tend to be smaller. 
In the numerical simulation given below, we use the maximum 
allowed values of $\phi_I$ constrained by Eq.~(\ref{phiinf}) 
as the initial condition of reheating.

\begin{figure}[h]
\begin{center}
\includegraphics[height=3.3in,width=3.4in]{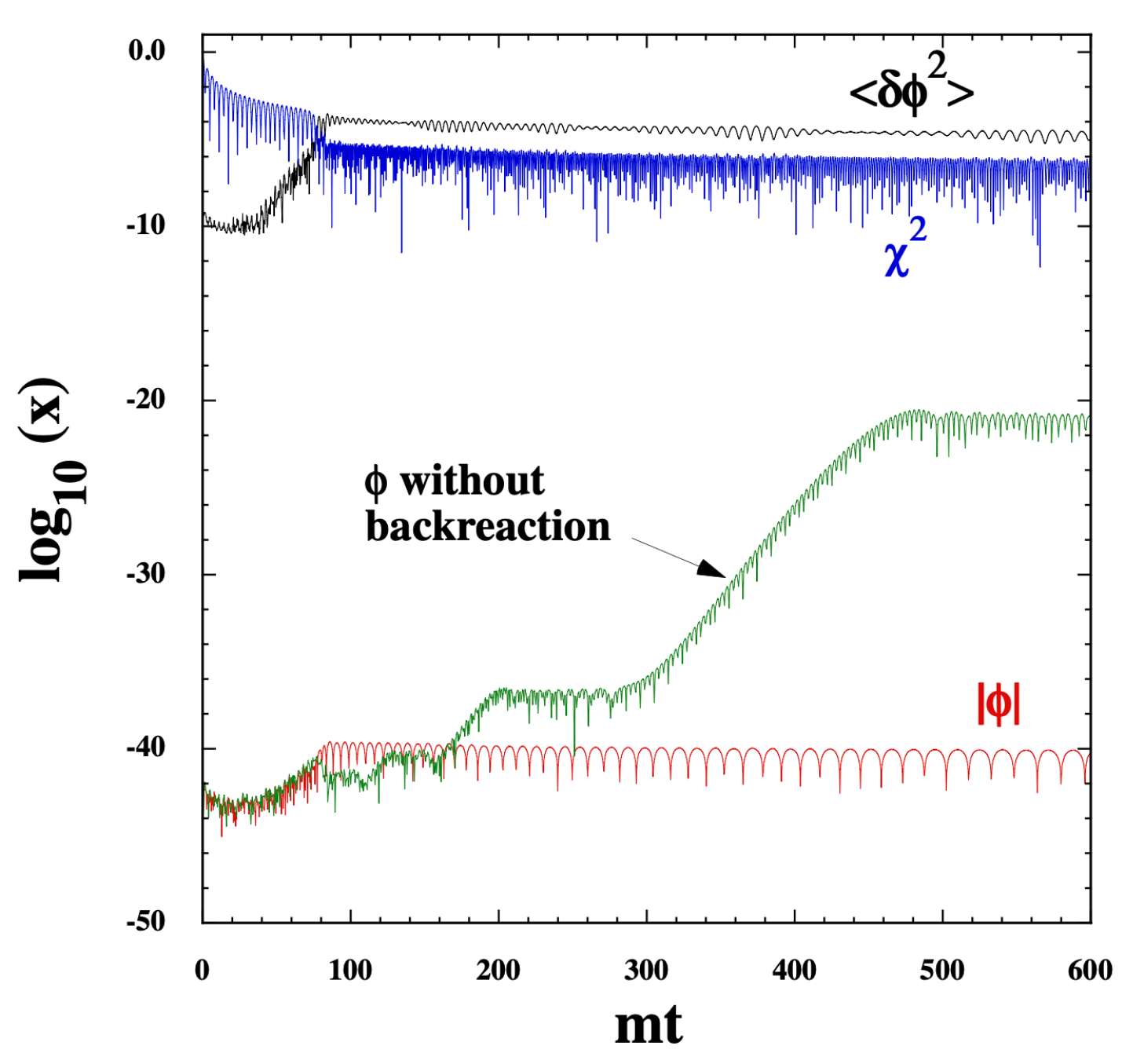}
\end{center}
\caption{\label{fig4}
Evolutions of $|\phi|,\chi^2,\langle \delta \phi^2 \rangle$
during the early stage of reheating for the quadratic 
potential $V(\chi)=m^2 \chi^2/2$ 
with $g=2.0 \times 10^{-3}$, 
$\beta=-4.4$, $m=6.0 \times 10^{-6}\Mpl$, and 
$\Gamma=1.0 \times 10^{-13}\Mpl$ 
($x=|\phi|,\chi^2,\langle \delta \phi^2 \rangle$ 
in the figure).
Here $\phi$ is normalized by 
$\Mpl$, while $\chi^2$ and $\langle \delta \phi^2 \rangle$ are 
normalized by $\Mpl^2$.
The initial field value $\phi_I$ at the onset of reheating 
is determined by the evolution of $\phi$ during inflation 
discussed in Sec.~\ref{infsec}. 
The red and green lines 
correspond to the cases in which the backreaction 
of created particles is included and neglected, 
respectively.
}
\end{figure}

In Fig.~\ref{fig4}, we show the evolutions of $|\phi|$, $\chi^2$, 
and $\langle \delta \phi^2 \rangle$ during the early stage of 
reheating for the coupling $g=2.0 \times 10^{-3}$ with $\beta=-4.4$, 
$m=6.0 \times 10^{-6}\Mpl$, and $\Gamma=1.0 \times 10^{-13}\Mpl$,
in which case $q_I=5.6 \times 10^4$. 
In this case, the variance $\langle \delta \phi^2 \rangle$ starts to
increase by parametric resonance and eventually catches up with the background inflaton density $\chi^2$. 
After $\langle \delta \phi^2 \rangle$ grows to the order 
$m^2/g^2$, the backreaction of created particles starts to violate 
the coherent oscillation of $\chi$. 
In Fig.~\ref{fig4}, this property can be seen after $\chi^2$ 
drops below $\langle \delta \phi^2 \rangle$. 
The growth of $\langle \delta \phi^2 \rangle$ terminates by 
the violation of coherent oscillations of $\chi$.
In the regime where $q_I$ is sufficiently larger than 1, 
our numerical analysis shows that the maximum values of 
$\langle \delta \phi^2 \rangle$ reached during preheating 
have the approximate dependence 
$\langle \delta \phi^2 \rangle_{\rm max} \propto 1/\sqrt{q_I}$.
This property agrees with what was found for a minimally coupled 
theory ($\beta=0$) \cite{Khlebnikov:1996wr}.
This means that, for increasing $g$ in the range $q_I \gg 1$, 
$\langle \delta \phi^2 \rangle_{\rm max}$ 
tends to be suppressed.

In Fig.~\ref{fig4}, we observe that the homogeneous field $\phi$ 
stops growing after $\langle \delta \phi^2 \rangle$ 
catches up with $\chi^2$. Hence the backreaction induced by 
the growth of $\langle \delta \phi^2 \rangle$ terminates the 
parametric excitation of $\phi$ as well. 
For $g=2.0 \times 10^{-3}$, the maximum value 
of $|\phi|$ reached during preheating is of order 
$\phi_{\rm max}=10^{-40}\Mpl$.
If we do not take the backreaction of created particles into account, 
$|\phi|$ continues to grow by the time 
at which the resonance parameter $q$ drops 
below the order 1. Without the backreaction effect, 
the maximum field value $\phi_{\rm max}$ for 
$g=2.0 \times 10^{-3}$ is of order $10^{-21}\Mpl$, 
which is enormously larger than the value 
$10^{-40}\Mpl$ obtained by implementing the 
backreaction (see Fig.~\ref{fig4}).

\begin{figure}[h]
\begin{center}
\includegraphics[height=3.1in,width=3.4in]{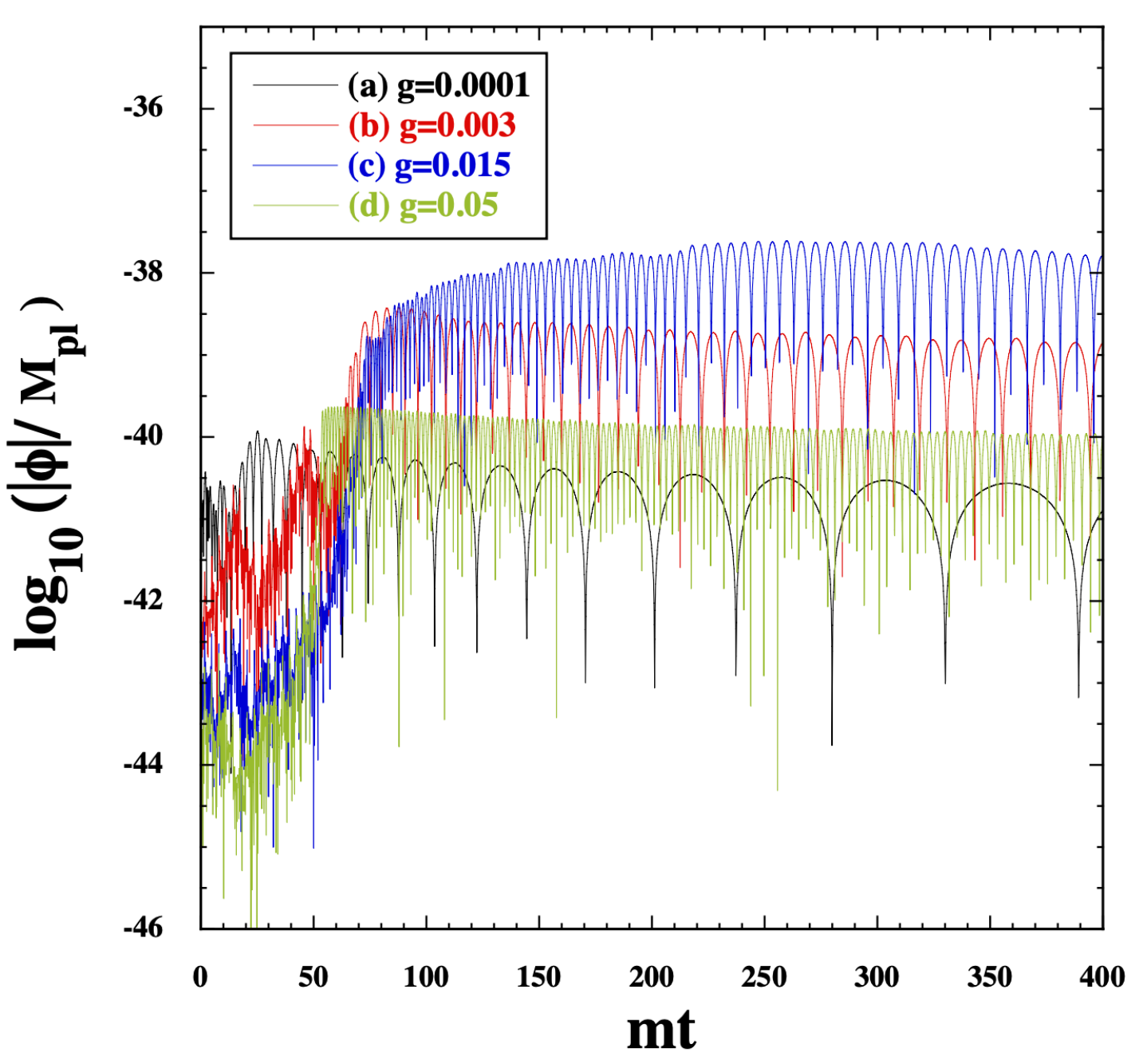}
\end{center}
\caption{\label{fig5}
Evolution of $|\phi|/\Mpl$ versus $mt$ during the early stage 
of reheating for the quadratic potential $V(\chi)=m^2 \chi^2/2$ 
with $\beta=-4.4$, $m=6.0 \times 10^{-6}\Mpl$,  
and $\Gamma=1.0 \times 10^{-13}\Mpl$. 
Each line corresponds to (a) $g=1.0 \times 10^{-4}$ (black), 
(b) $g=3.0 \times 10^{-3}$ (red), 
(c) $g=1.5 \times 10^{-2}$ (blue), and 
(d) $g=5.0 \times 10^{-2}$ (green). }
\end{figure}

In Fig.~\ref{fig5}, we plot the evolution of $|\phi|$ for four different 
values of $g$. Provided that $g \gtrsim 10^{-4}$, $|\phi|$ is 
initially amplified by parametric resonance.
For $g=1.0 \times 10^{-4}$, which corresponds to the parameter 
$q_I=1.4 \times 10^2$, there is an initial short period in which 
$|\phi|$ is enhanced, but the growth is limited due to the 
early entry to the region $q \lesssim 1$. 
For $g \gtrsim 10^{-2}$, the growth of $|\phi|$ is 
terminated by the backreaction of created particles. 
When $g \simeq 10^{-2}$, we find that 
$\phi_{\rm max}$ is of order $10^{-38}\Mpl$. 
In this case, the maximum value of $|\phi|$ obtained 
without the backreaction is of order $10^{-3}\Mpl$, 
which is $10^{35}$ times as large as the value $10^{-38}\Mpl$.
This shows the importance of properly implementing the 
backreaction effect to estimate the maximum value of 
$\phi_{\rm max}$ reached during preheating. 

As $g$ increases in the coupling range $g \gtrsim 10^{-2}$, 
$\phi_{\rm max}$ tends to be suppressed in comparison to 
the value around $g=10^{-2}$. 
Indeed, we can confirm this property 
in Fig.~\ref{fig5} for
the coupling $g=5.0 \times 10^{-2}$.
Then, for $g \gtrsim 10^{-4}$, the maximum values of 
$|\phi|$ obtained under the Hartree approximation are in the range
\be
\phi_{\rm max} \lesssim 10^{-38}\Mpl\,.
\ee
This is significantly smaller than the upper limit  
$|\phi_{\rm R}| \simeq 10^{-11}\Mpl$ constrained from the post-Newtonian bound at the end of reheating.

We note that our approximation scheme does not incorporate 
nonlinear effects like rescattering. 
However, it is known that the rescattering of $\phi$ particles off 
the inflaton condensate tends to limit the growth of 
$\langle \delta \phi^2 \rangle$ further \cite{Khlebnikov:1996zt}.
This means that the growth of the homogeneous field $\phi$ should 
be also limited, so it is expected that the maximum values of $\phi$ 
do not exceed those derived under the Hartree approximation. 
After the rescattering of produced particles, 
the variance $\langle \delta \phi^2 \rangle$ 
reaches an equilibrium state.
At this stage,  
the significant amplification of the homogeneous field from 
$|\phi| \lesssim 10^{-38}\Mpl$ to the value exceeding 
the order $10^{-11}\Mpl$ by the end of reheating 
is unexpected. 

In particular, the effective mass squared (\ref{meff}) 
is approximately given by $m_{\rm eff}^2 \simeq g^2 \chi^2+3\beta H^2/2$ 
in the temporal matter era during reheating. 
Until the onset of the radiation era at which the inflaton field $\chi$ 
completely decays to radiation, the negative coupling term $3\beta H^2/2$ 
does not completely dominate over the positive contribution 
$g^2 \chi^2$ to $m_{\rm eff}^2$. 
Hence the strong tachyonic instability of $\phi$ 
induced by the negative nonminimal coupling constant $\beta$  
is not expected at the late stage of reheating. 
Eventually, the Born decay term 
$\Gamma \dot{\chi}$ in Eq.~(\ref{chieq}) starts to work 
to convert the inflaton density to the radiation density. 

Thus, for $g \gtrsim 10^{-4}$, the growth of $|\phi|$ saturated by 
the backreaction during preheating allows the field value 
$\phi_{\rm R}$ at the end of reheating consistent with the post-Newtonian 
constraint.

\begin{figure}[h]
\begin{center}
\includegraphics[height=3.2in,width=3.5in]{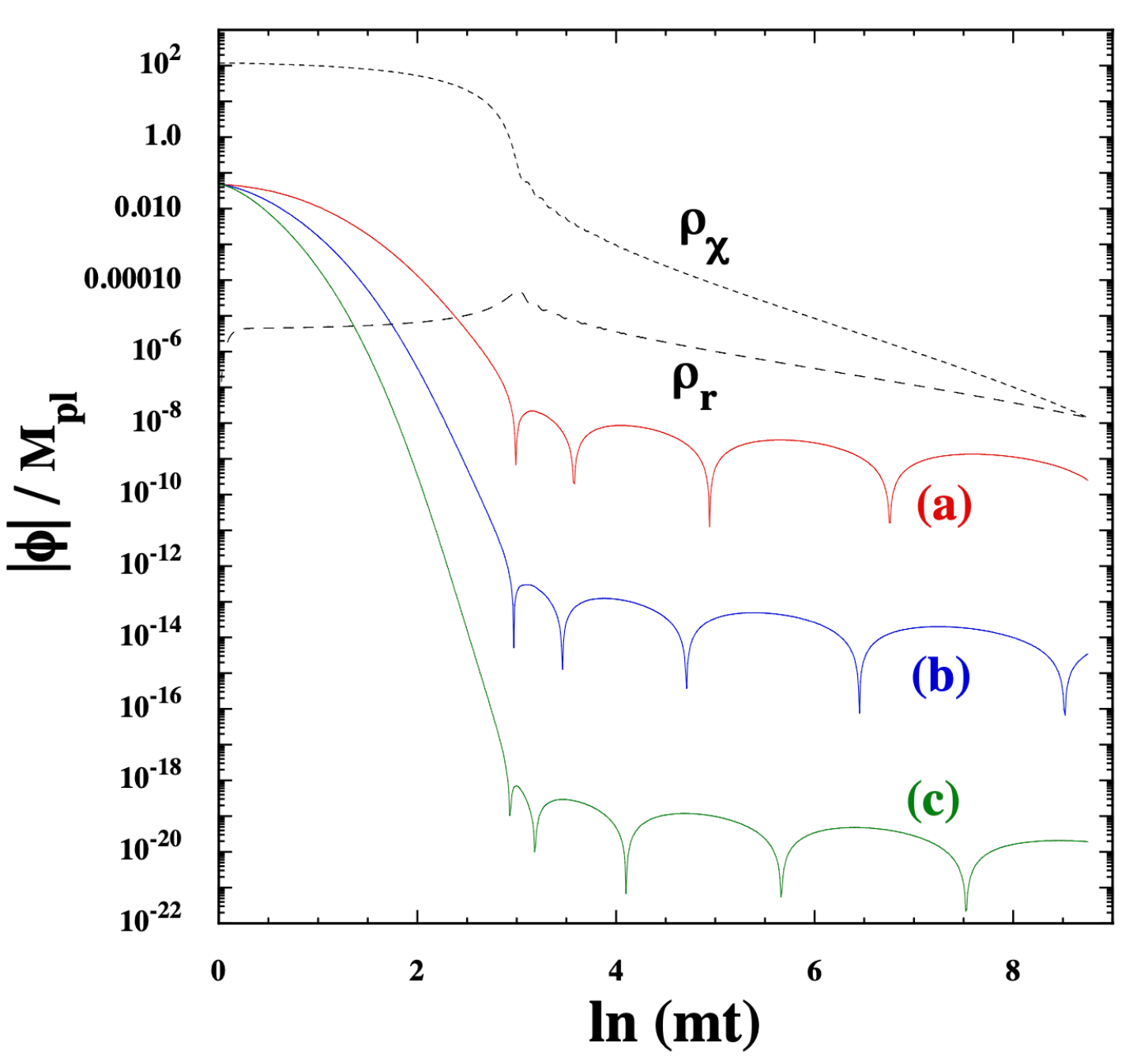}
\end{center}
\caption{\label{fig6}
Evolution of $|\phi|/\Mpl$ versus $\ln (mt)$ during inflation and 
reheating for the quadratic potential $V(\chi)=m^2 \chi^2/2$ 
with $\beta=-4.4$, $m=6.0 \times 10^{-6}\Mpl$, and $\Gamma=1.0 \times 10^{-9}\Mpl$. 
Each line corresponds to 
(a) $g=1.27 \times 10^{-5}$ (red), 
(b) $g=1.28 \times 10^{-5}$ (blue), and 
(c) $g=1.29 \times 10^{-5}$ (green). 
Two black dashed lines show the evolution of 
the inflaton density $\rho_{\chi}$ and radiation density 
$\rho_r$. We integrate the background equations by the time 
at which $\rho_r$ catches up with $\rho_{\chi}$.
}
\end{figure}

For $g \lesssim 10^{-4}$, the preheating stage is absent and hence 
we do not need to incorporate the backreaction effect. 
For $g \lesssim 10^{-5}$, the coupling $g$ does not overwhelm 
the negative nonminimal coupling $\beta$ in the effective 
mass of $\phi$. 
Then, $\phi$ is not subject to the 
strong suppression during inflation. 
In Fig.~\ref{fig6}, we plot the evolution of $|\phi|/\Mpl$ 
during inflation and reheating for three different values 
of $g$ with the decay constant $\Gamma=1.0 \times 10^{-9}\Mpl$.
We integrate the background equations of motion by the end 
of reheating (time $t_{\rm R}$) at which $\rho_r$ 
catches up with the inflaton density 
$\rho_{\chi}=\dot{\chi}^2/2+V(\chi)$. 
In case (a) the field value at $t=t_{\rm R}$ is of order $10^{-10}\Mpl$, 
so it exceeds the solar-system bound 
$|\phi_{\rm R}| \lesssim 10^{-11}\Mpl$.
In cases (b) and (c), the field values at $t=t_{\rm R}$ 
are of orders $10^{-15}\Mpl$ and $10^{-20}\Mpl$, respectively, 
which are well within the bound 
$|\phi_{\rm R}| \lesssim 10^{-11}\Mpl$. 
This means that, for $\Gamma=1.0 \times 10^{-9}\Mpl$, 
the coupling in the range $g \geq 1.28 \times 10^{-5}$ can be consistent with the solar-system constraint. 

In Fig.~\ref{fig6}, we observe that, after the initial rapid decrease of 
$|\phi|$ during inflation, $|\phi|$ exhibits mild decrease in the reheating period. 
This reflects the fact that $g^2 \chi^2$ is larger than the term 
$3|\beta|H^2/2$ during reheating. 
The field value $\phi_{\rm R}$ at the end of reheating depends on 
the decay constant $\Gamma$. For decreasing 
$\Gamma$, $|\phi_{\rm R}|$ tends to be smaller 
because of the longer period of reheating. 
We also note that the field value $\phi_{I}$ 
at the beginning of reheating 
depends on the duration of inflation. 
For the minimum number of e-foldings $N=60$, 
the criterion consistent with the bound 
$|\phi_{\rm R}| \lesssim 10^{-11}\Mpl$ is given by 
\be
g \geq 1.3 \times 10^{-5}\,,
\ee
irrespective of the values of $\Gamma$ smaller than $H$ 
at the end of inflation.
Thus, the mechanism proposed by Anson {\it et al.} \cite{Anson:2019uto} 
works even for small couplings of order $10^{-5}$.

\subsection{$\alpha$-attractor with $\alpha=1$}

\begin{figure}[h]
\begin{center}
\includegraphics[height=3.1in,width=3.4in]{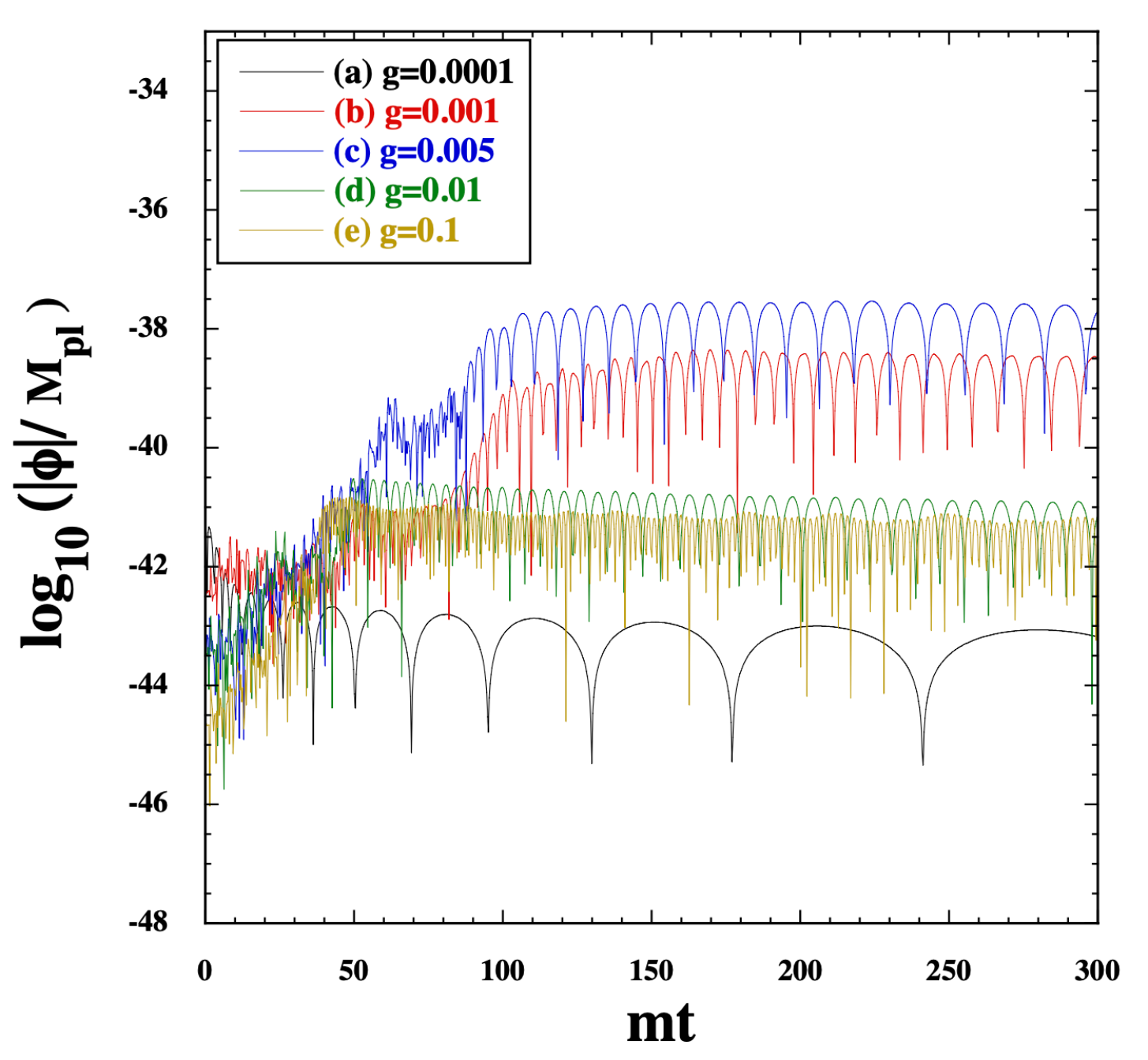}
\end{center}
\caption{\label{fig7}
Evolution of $|\phi|/\Mpl$ versus $mt$ during the early stage 
of reheating for the $\alpha$-attractor potential (\ref{Vchi})
with $\alpha=1$, $\beta=-4.4$, $m=1.1 \times 10^{-5}\Mpl$, 
and $\Gamma=1.0 \times 10^{-13}\Mpl$. 
Each line corresponds to (a) $g=1.0 \times 10^{-4}$ (black), 
(b) $g=1.0 \times 10^{-3}$ (red), (c) $g=5.0 \times 10^{-3}$ (blue), 
(d) $g=1.0 \times 10^{-2}$ (green), and 
(e) $g=1.0 \times 10^{-1}$ (brown). }
\end{figure}

We also study the dynamics of $\phi$ in the $\alpha$-attractor 
model with $\alpha=1$. 
In the $\alpha$-attractor model, the field value 
$\chi_I=0.940 \Mpl$ at the onset 
of reheating is smaller than that for the exact quadratic potential. 
Hence we require larger couplings $g$ to enhance 
both $\phi$ and $\langle \delta \phi^2 \rangle$ 
by parametric resonance in comparison to the case of 
quadratic potential. 
In the numerical simulation of Fig.~\ref{fig7},
we observe that $|\phi|$ does not grow for $g=1.0 \times 10^{-4}$, 
but parametric resonance occurs for $g=1.0 \times 10^{-3}$.

Numerically, we find that the maximum value of $|\phi|$ reached 
for $g=5.0 \times 10^{-3}$ is of order $\phi_{\rm max} =10^{-38} \Mpl$ 
under the Hartree approximation. 
As $g$ increases in the range $g \gtrsim 5 \times 10^{-3}$, 
$\phi_{\rm max}$ tends to be decreased. 
This property can be seen in Fig.~\ref{fig7} for the 
couplings $g=1.0 \times 10^{-2}$ and $g=1.0 \times 10^{-1}$. 
Thus, for any couplings with $g \gtrsim 10^{-4}$, 
the maximum values of $|\phi|$ reached during preheating 
are in the range 
\be
\phi_{\rm max} \lesssim 10^{-38}\Mpl\,,
\label{phimax2}
\ee
which are again much smaller than the post-Newtonian upper limit 
$\phi_{\rm R} \simeq 10^{-11}\Mpl$ at the end of reheating. 

If we ignore the backreaction of created $\phi$ particles, 
$\phi_{\rm max}$ can be significantly 
larger than the upper limit (\ref{phimax2}).
When $g=1.0 \times 10^{-1}$, for example, the numerical value 
of $\phi_{\rm max}$ derived by neglecting the backreaction effect
is of order $10^{-2}\Mpl$, which is very much larger 
than $10^{-11}\Mpl$. 
Thus, inclusion of the backreaction is crucially important 
for the proper estimation of $\phi_{\rm max}$.
The upper limit $\phi_{\rm max}={\cal O}(10^{-38} \Mpl)$ 
derived in the presence of the preheating stage 
is similar to that obtained for the quadratic inflaton potential.

\begin{figure}[h]
\begin{center}
\includegraphics[height=3.2in,width=3.5in]{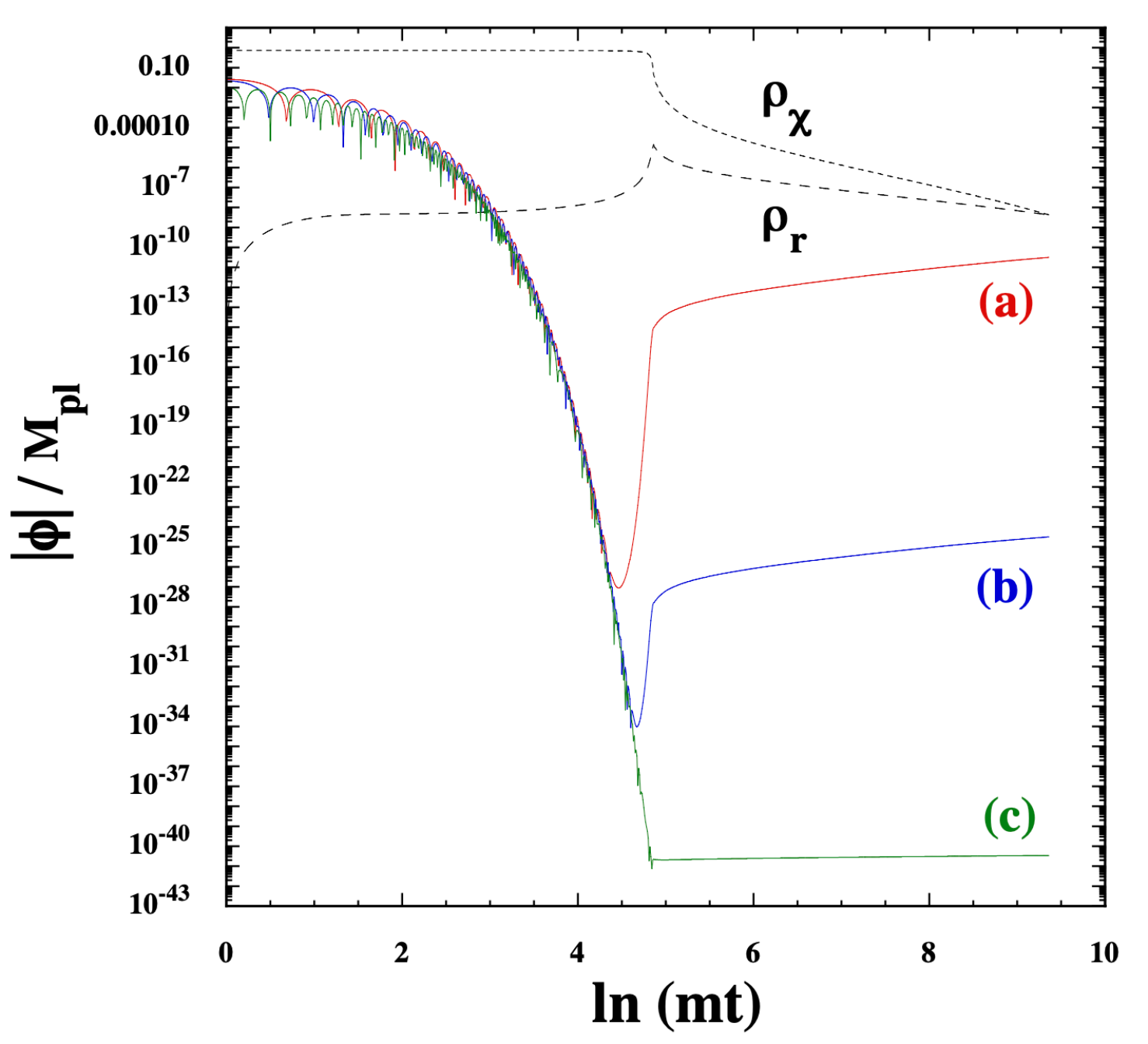}
\end{center}
\caption{\label{fig8}
Evolution of $|\phi|/\Mpl$ versus $\ln (mt)$ during inflation 
and reheating for the $\alpha$-attractor potential (\ref{Vchi})
with $\alpha=1$, $\beta=-4.4$, $m=1.1 \times 10^{-5}\Mpl$, and 
$\Gamma=1.0 \times 10^{-9}\Mpl$. 
Each line corresponds to 
(a) $g=6.7 \times 10^{-6}$ (red), 
(b) $g=8.0 \times 10^{-6}$ (blue), and 
(c) $g=1.6 \times 10^{-5}$ (green). 
Two black dashed lines show the evolution of 
the inflaton density $\rho_{\chi}$ and radiation density 
$\rho_r$. The background equations are integrated 
by the time at which $\rho_r$ catches up with $\rho_{\chi}$.
}
\end{figure}

For the coupling $g$ smaller than the order $10^{-5}$, 
the field $\phi$ is not subject to strong suppression 
during inflation. Since parametric resonance does not 
occur for such small couplings, we do not need to 
implement the backreaction of created $\phi$ particles. 
In Fig.~\ref{fig8}, we show the evolution of $|\phi|$ 
during inflation and reheating for three different 
values of $g$ with the decay constant $\Gamma=1.0 \times 10^{-9}\Mpl$. 
In cases (a) and (b) the field values 
at the end of reheating (time $t=t_{\rm R}$) 
are of order $10^{-11}\Mpl$ and $10^{-25}\Mpl$, respectively,
so case (b) is consistent with the solar system limit 
$|\phi_{\rm R}| \lesssim 10^{-11} \Mpl$. 

We need to caution that, in both cases (a) and (b), $|\phi|$ grows 
during reheating due to the dominance of the negative nonminimal coupling 
$3 \beta H^2/2$ relative to $g^2 \chi^2$ in $m_{\rm eff}^2$ 
(whose dominance starts to occur during inflation).
If we consider a very low energy-scale reheating 
where the reheating temperature 
is of order MeV \cite{Hannestad:2004px,Hasegawa:2019jsa}, the decay 
constant is of order $\Gamma \simeq 1~{\rm sec}^{-1} \simeq 10^{-43}\Mpl$. 
In this case, the time at the end of reheating is estimated as 
$t_{\rm R} \simeq 1/\Gamma \simeq 10^{43}\Mpl^{-1}$ and hence 
$\ln (mt_{\rm R}) \simeq 88$. 
In case (b), for example, the order of $|\phi|$ increases by 
one order of magnitude during the time interval 
$\ln (m \Delta t)=2$. Then, for the MeV scale reheating, 
the field value $\phi_{\rm R}$ exceeds the limit even 
in case (b). 

To avoid the increase of $|\phi|$ during reheating, 
we require that $m_{\rm eff}^2$ is positive by the end of inflation. 
As we discussed in Sec.~\ref{alinfs}, this translates to the 
condition $g>1.61 \times 10^{-5}$ for the $\alpha$-attractor 
with $\alpha=1$. 
In case (c) of Fig.~\ref{fig8}, we plot the evolution of 
$|\phi|$ for $g=1.6 \times 10^{-5}$. 
This corresponds to the 
marginal case in which the monotonic growth of $|\phi|$ during 
reheating can be avoided.  
Then, provided that 
\be
g \geq 1.6 \times 10^{-5}\,,
\ee
the field value at the end of reheating does not exceed the 
solar-system limit irrespective of the decay constant $\Gamma$. 
For the coupling range $g \gtrsim 10^{-3}$ in which the preheating 
epoch is present, 
the maximum field value is limited as Eq.~(\ref{phimax2}). 
In this coupling regime, $g^2 \chi^2$ dominates over
$3|\beta|H^2/2$ after inflation and hence the growth 
of $|\phi|$ from the end of preheating to the onset of radiation 
era is not expected to occur.

\section{Conclusions}
\label{consec}

In this paper, we studied the cosmological evolution of a scalar field $\phi$ 
in the presence of a nonminimal coupling $e^{-\beta \phi^2/(2M_{\rm pl}^2)}R$
and a four-point coupling $g^2 \phi^2 \chi^2/2$ between $\phi$ and the 
inflaton field $\chi$. 
This nonminimal coupling gives rise to the phenomenon 
of spontaneous scalarization of NSs for $\beta \leq -4.35$. 
If we apply the original DEF scenario to cosmology, the scalar field $\phi$ 
is subject to tachyonic instability during 
the periods of cosmic acceleration and matter dominance. 
In Ref.~\cite{Anson:2019ebp}, it was argued that 
the coupling $g^2 \phi^2 \chi^2/2$ allows a possibility 
for curing this problem by realizing a positive mass squared during inflation.

Since the dynamics of the field $\phi$ during the post inflationary reheating 
period was not addressed in the literature, we have studied the cosmological 
evolution from the onset of inflation to today including the reheating stage.
To satisfy solar-system constraints, the field value at the beginning of  
radiation era is constrained as Eq.~(\ref{phiR2}). 
For the coupling $\beta$ relevant to the occurrence of 
spontaneous scalarization, this bound corresponds to
$|\phi_{\rm R}| \lesssim 10^{-11}\Mpl$.

Provided that the effective mass squared (\ref{meff}) 
is larger than the order of $H^2$ during inflation, 
the field $\phi$ is subject to the exponential suppression 
($|\phi| \propto a^{-3/2}$) by the end of inflation. 
For the quadratic and $\alpha$-attractor potentials with $\alpha=1$, 
we showed that this suppression of $\phi$ occurs for the coupling $g$ 
in the ranges (\ref{gconm}) and (\ref{gcona}), respectively, 
whose minimum values are both of order $10^{-5}$. 
With these two potentials, the parametric excitation of $\phi$ and its perturbations during preheating can occur for $g \gtrsim 10^{-4}$ and $g \gtrsim 10^{-3}$, respectively.
If we do not take the backreaction of created $\phi$ particles into account, 
the maximum values of $|\phi|$ reached during preheating can exceed 
the order of $10^{-11}\Mpl$. Incorporating the backreaction under the Hartree 
approximation, however, we found that $\phi_{\rm max}$ is smaller than the 
order of $10^{-38}\Mpl$. After the termination of parametric resonance, 
the further growth of $|\phi|$ is unexpected by the end of reheating 
because the $g^2 \chi^2$ term dominates over the negative nonminimal coupling 
in the equation of motion of $\phi$.

In the regime of small couplings $g$ without the preheating stage, 
we also numerically solved the background equations of motion by the end of 
reheating with the Born decay term taken into account. 
For the quadratic inflaton potential, the amplitude of $\phi$ decreases 
during reheating in the range of couplings $g$ that gives 
a positive mass squared during inflation. 
In this case, provided $g \geq 1.3 \times 10^{-5}$, the model is consistent 
with the solar-system bound $|\phi_{\rm R}| \lesssim 10^{-11}\Mpl$.
For the $\alpha$-attractor with $\alpha=1$, $m_{\rm eff}^2$ can 
change its sign during inflation by the presence of a negative nonminimal coupling
even if $m_{\rm eff}^2>0$ at the onset of inflation. 
In such cases, the growth of $|\phi|$ also occurs in the reheating period, 
see Fig.~\ref{fig8}. If we consider a low-scale reheating scenario with the
reheating temperature of order MeV, the monotonic growth of $|\phi|$ 
during a long period of the reheating era can 
conflict with the limit $|\phi_{\rm R}| \lesssim 10^{-11}\Mpl$.
Provided $g \geq 1.6 \times 10^{-5}$, we found that  
the $\alpha$-attractor with $\alpha=1$ can be consistent with 
the solar-system bound irrespective of the decay constant $\Gamma$.

We thus showed that, for natural couplings in the range $g \gtrsim 10^{-5}$, 
the scenario proposed by Anson {\it et al}. leads to the viable 
cosmological evolution of $\phi$ consistent with today's local 
gravity constraints. Since the inflaton field decays to radiation 
by the onset of radiation era, it does not affect the process of 
spontaneous scalarization of NSs which can occur in later 
cosmological epochs. 
Interestingly, the field $\phi$ responsible 
for spontaneous scalarization can also exhibit the 
phenomenon of parametric resonance in the early Universe.
It is then possible to probe this scenario not only from 
the gravitational waveform emitted from compact binaries  
but also from the gravitational wave background.

While we focused on the cosmology in the DEF model with the 
coupling $g^2 \phi^2 \chi^2/2$, it may be of interest to 
extend the analysis to the cases in which higher-order $\phi$-dependent terms 
like $\phi^4$ in $F(\phi)$ \cite{Anderson:2016aoi} 
or k-essence terms like $X^2$ \cite{Higashino:2022izi} are also present.
This may widen parameter spaces of the coupling constant $\beta$ 
consistent with binary pulsar constraints. 
These issues are left for future works.

\section*{ACKNOWLEDGMENTS}
We thank Masato Minamitsuji 
for useful discussions. 
This research project is supported by National Research Council of 
Thailand (NRCT): NRCT5-RGJ63009-110.
SP is supported by a Waseda University Grant for Special Research 
Project (No.~2022C- 632). 
ST is supported by the Grant-in-Aid for 
Scientific Research Fund of the JSPS Nos.~19K03854 and 22K03642.


\bibliographystyle{mybibstyle}
\bibliography{bib}

\end{document}